\documentclass[aps,pra,superscriptaddress,twocolumn]{revtex4-1}

\usepackage{amssymb}
\usepackage{graphicx}
\usepackage{amsmath}
\usepackage{amsthm}
\usepackage{amsfonts}
\usepackage{color}
\usepackage{bm}
\usepackage{bbm}
\usepackage{url}
\usepackage{amssymb}
\usepackage[driverfallback=dvipdfm]{hyperref}
\hypersetup{pdfpagemode=FullScreen,colorlinks=true,breaklinks,urlcolor=blue,linkcolor=blue,citecolor=blue}

\usepackage{txfontsb}
\usepackage[OT1]{fontenc}
\usepackage{soul}

\begin{document}

\makeatletter
\newcommand{\Rmnum}[1]{\expandafter\@slowromancap\romannumeral #1@}
\makeatother

\global\long\def\id{\mathbbm{1}}
\global\long\def\ui{\mathbbm{i}}
\global\long\def\ud{\mathrm{d}}

\title{Characterizing Floquet topological phases by quench dynamics: A multiple-subsystem approach}

\author{Bei-Bei Wang}
\affiliation{School of Physics and Institute for Quantum Science and Engineering, Huazhong University of Science and Technology, Wuhan 430074, China}

\author{Long Zhang}
\email{lzhangphys@hust.edu.cn}
\affiliation{School of Physics and Institute for Quantum Science and Engineering, Huazhong University of Science and Technology, Wuhan 430074, China}
\affiliation{Hefei National Laboratory, Hefei 230088, China}

\begin{abstract} 
We investigate the dynamical characterization theory for periodically driven systems in which Floquet topology can be fully detected by emergent topological patterns of quench dynamics in momentum subspaces called band-inversion surfaces.
We improve the results of a recent work [Zhang {\em et al.},Phys. Rev. Lett. {\bf 125}, 183001 (2020)] and propose a more flexible scheme 
to characterize a generic class of $d$-dimensional Floquet topological phases classified by $\mathbb{Z}$-valued invariants
by applying a quench along an arbitrary spin-polarization axis.
Our basic idea is that by disassembling the Floquet system into multiple static subsystems that are periodic in quasienergy,
a full characterization of Floquet topological phases reduces to identifying a series of bulk topological invariants for time-independent Hamiltonians, 
which greatly enhances the convenience and flexibility of the measurement.
We illustrate the scheme by numerically analyzing two experimentally realizable models in two and three dimensions, respectively,
and adopting two different but equivalent viewpoints to examine the dynamical characterization.
Finally, considering the imperfection of experiment, we demonstrate that the present scheme 
can also be applied to a general situation where the initial state is not completely polarized.
This study provides an immediately implementable approach for dynamically classifying Floquet topological phases in ultracold atoms or other quantum simulators.

\end{abstract}

\maketitle

\section{Introduction}

Topological quantum phases have attracted extensive exploration in the last decades~\cite{TI_review1,TI_review2,Sato2017,Fang2016}. 
Compared with solid materials, ultracold atoms provide a clean platform with super controllability and stability for simulating exotic topological phases~\cite{Zhang2018_review,Cooper2019_review,Schafer2020_review}. 
Recent significant experimental advances include the realization of a one-dimensional (1D) Su-Schrieffer-Heeger chain~\cite{Atala2013} and chiral topological phase~\cite{Song2018}, two-dimensional (2D) Chern insulators \cite{Aidelsburger2013,Miyake2013,Jotzu2014,Aidelsburger2015,Wu2016,Sun2018a,Liang2023} and three-dimensional (3D) topological semimetals~\cite{Song2019,Wang2021}. 
In parallel, much theoretical progress has also been made~\cite{Liu2013,Liu2014,Zhou2017,WangBZ2018,LuWang2020,BBWang2021,Ziegler2022}.

Floquet engineering has been a versatile tool for tailoring quantum phases~\cite{Eckardt2017_review,Harper2020_review,Rudner2020_review,Weitenberg2021_review,Rudner2013,Goldman2014,Bukov2015,Hockendorf2019,Liu2019,Peng2019,Hu2020a,Huang2020,Jangjan2022,Unal2022,Li2023}.
With coherent temporal control, a periodically driven system can be engineered not only to 
simulate an effective static Hamiltonian such as the Haldane model~\cite{Jotzu2014},
but also to realize novel topological phases without static counterparts~\cite{Wintersperger2020}. 
A typical example is the anomalous Floquet topological phase, whose boundary modes exist in two quasienergy gaps 
and have no direct correspondence to the bulk topology~\cite{Rudner2013}. 
Such a $d$-dimensional ($d$D) anomalous Floquet topological phase is conventionally characterized by topological invariants defined in the full
$(d+1)$D space-time dimension~\cite{Rudner2013,Kitagawa2010,Nathan2015,Carpentier2015,Fruchar2016,Roy2017,Yao2017,Morimoto2017,Unal2019}, 
which are, however, not convenient for experimental measurements.  
To overcome this issue, a much simplified characterization theory has been recently developed in which the complex evolution in time domain need not be considered,
and subdimensional topology defined in particular $(d-1)$D momentum subspaces called band-inversion surfaces (BISs) is introduced to fully characterize Floquet topological phases~\cite{Zhang2020}. 
This result indicates a precise and systematic route for quantum engineering of unconventional Floquet topological phases~\cite{Zhang2022}, 
and has enabled the experimental realization of highly tunable anomalous Floquet topological bands~\cite{Zhang2023}.


An important advantage of the BIS characterization is that the topology on lower-dimensional BISs can be directly detected by quench dynamics~\cite{Zhang2018,Zhang2019a,Zhang2019b,Hu2020b,Ye2020,Yu2021,Jia2021,LiZhuGong2021,Fang2022,He2023}, particularly since quantum quenches are easy to implement in ultracold atoms~\cite{Song2018,Song2019,Wang2021,Langen2015,Caio2015,Flaschner2016,Wang2017,Flaschner2018,Tarnowski2019,Sun2018b,Yi2019,Unal2020}. 
For a static $d$D topological system, it has been demonstrated that the BISs are marked by quench-induced resonant spin oscillations, 
and the subdimensional topological invariant that classifies the $d$D bulk topology can be measured by dynamical topological patterns emerging on all $(d-1)$D BISs~\cite{Zhang2018,Zhang2019a,Zhang2019b}.
This renders the dynamical bulk-surface correspondence in the momentum space, which plays a similar role to that of the conventional bulk-boundary correspondence for equilibrium topological phases in the real space, and has brought much experimental progress in the simulation and characterization of topological phases using ultracold atoms~\cite{Sun2018b,Yi2019} or other quantum simulators~\cite{Wang2019,Ji2020,Xin2020,Yu2021,Niu2021}.
In Ref.~\cite{Zhang2020}, built on a generalized dynamical bulk-surface correspondence, 
a dynamical scheme is proposed to realize the BIS characterization of generic $d$D Floquet systems. 
Very recently, identifying the Floquet topology through dynamically detecting the BIS configuration has been applied in cold-atom experiments~\cite{Zhang2023}.
Therefore, a comprehensive and detailed study of how to characterize Floquet topological phases by quench dynamics is necessary and timely.

However, we find that the generalization of the dynamical characterization scheme from static systems to periodically driven systems is not fully straightforward.
The key difference is that due to the existence of two quasienergy gaps, one needs to choose a proper spin-polarization axis to define the BISs of the Floquet bands,
such that the subdimensional topology defined on BISs can completely characterize the topology of both quasienergy gaps. 
In contrast, the BIS characterization of a static topological phase does not have a special axis.
This difference means that the direct generalization of dynamical characterization cannot be as flexible as that for static systems.
For example, the proposed dynamical scheme in Ref.~\cite{Zhang2020} works under the condition that the quenched spin-polarization axis 
is exactly the one that defines the BIS, which may limit implementation in real experiments.

In this paper, we improve the results in Ref.~\cite{Zhang2020} and propose a more flexible scheme 
that can fully characterize a generic class of $d$D $\mathbb{Z}$-invariant Floquet topological phases 
by quenching an arbitrary spin-polarization axis.
These Floquet topological phases may fall into the Altland-Zirnbauer (AZ) symmetry classes characterized by $\mathbb{Z}$ topological invariants~\cite{Roy2017,Yao2017}, 
and can also be beyond the conventional classification by global bulk topology~\cite{Zhang2022}.
Starting from an initial static phase that is fully polarized in one direction,
the quench is realized by instantaneously decreasing the Zeeman field in the polarized axis and, 
at the same time, turning on the periodic driving [see Fig.~\ref{Fig1}(c)].
The main idea of the present approach is that a Floquet system can be disassembled into multiple static subsystems, 
so that its dynamical characterization is turned into a characterization of one or more time-independent bulk Hamiltonians.
Here each subsystem consists of the bands that are inverted on a BIS within either quasienergy gap.
In this way, the method of dynamically classifying static topological phases can be applied and it is independent of the choice of the quench axis. 
We illustrate this multiple-subsystem approach with 2D and 3D periodically driven models. 
Both of the two models are experimentally feasible with ultracold atoms~\cite{Wu2016,Sun2018a,Liang2023,Zhang2023} or solid-state spin systems~\cite{Ji2020,Xin2020}.
In the characterization of 2D anomalous Floquet topological phases, we consider quantum quenches along different spin-polarization axes 
and construct topological invariants from two equivalent viewpoints: (i) the winding of an emergent dynamical spin-texture field 
on {\it dynamical} band-inversion surfaces (dBISs) and (ii) the total charges of the dynamical field enclosed by dBISs.
Here, the dBISs are the momentum subspaces that are identified by quench dynamics to define the subdimensional topology.
It should be noted that the dBIS is equal to a BIS of the Floquet bands only when the quench axis is exactly the one defining the BIS, 
but the dynamical topology emerging on dBISs is the same for all quench ways.
As an important supplement and extension, we also discuss a more general situation 
where the quench starts from an initial state that is not fully polarized (dubbed a `` shallow quench'').
Such discussion of shallow quenches improves the applicability of our dynamical scheme.

This paper is organized as follows. In Sec.~\ref{secII}, we review the basic idea underlying the BIS characterization theory for both static and Floquet systems. 
Several key concepts are introduced. In Sec.~\ref{secIII}, we first propose a generic scheme and examine two cases: The quench axis is equal to or not equal to the one defining the BIS. 
We then illustrate the scheme by numerically calculating two highly feasible models in two and three dimensions, respectively. 
In Sec.~\ref{secIV}, we adopt the viewpoint of topological charges to demonstrate again the dynamical characterization.
The 2D driven model serves as an illustrative example.
In Sec.~\ref{secV}, we discuss shallow quenches and demonstrate that the dynamical scheme can work widely.
A brief discussion and summary are presented in Sec.~\ref{secVI}. More details are given in the Appendixes.

%

\section{Model and concepts}~\label{secII}

We consider a class of periodically driven systems described by the Hamiltonian
\begin{equation}~\label{Ham_total}
H({\bf k}, t) = H_s(\bold{k})+V(\bold{k},t),\quad H_s(\bold{k})=\sum_{i=0}^{d}h_{i}(\bold{k})\gamma_{i}
\end{equation}
where $H_s(\bold{k})=\bold{h}(\bold{k})\cdot \boldsymbol{\gamma}$ represents a $d$D gapped topological phase (insulator or superconductor) characterized by integer invariants 
in the AZ symmetry classes~\cite{AZ1997,Schnyder2008,Kitaev2009,Chiu2016} (see Appendix~\ref{App1}), 
and $V(\bold{k},t)$ is a periodic drive that can take a general form $V({\bf k},t)=V_{l_1}({\bf k},t)\gamma_{l_1}+V_{l_2}({\bf k},t)\gamma_{l_2}+\cdots$.
Here, $l_i\in\{0,1,\cdots,d\}$ and $V_{l_i}(t)=V_{l_i}(t+T)$ with $T$ being the driving period. 
The $\gamma$ matrices obey the Clifford algebra $\{\gamma_i,\gamma_j\}=2 \delta_{ij}\id$ ($i,j=0,1,\dots,d$) and the $(d + 1)$D vector $\bold{h}(\bold{k})$ depends on the momentum $\bold{k}$ in the first Brillouin zone (BZ). 
We emphasize that the $\gamma$ matrices are arranged in an order that satisfies the trace property~\cite{Zhang2018}
\begin{align}~\label{trace}
{\rm Tr}\left(S\prod_{i=0}^{d}\gamma_i\right)&=(-2\ui)^n,
\end{align}
where $S=\ui^n\prod_{i=0}^{d}\gamma_i$ is the chiral matrix (or the identity matrix $\id$) of dimension $n_d = 2^n$ for odd $d=2n-1$ (even $d=2n$). 
For odd dimensions, the topological phase requires chiral-symmetry protection, ensured by a restriction $V(t_{\rm ref}+t)=V(t_{\rm ref}-t)$ ($0\leqslant t_{\rm ref}<T$)~\cite{Zhang2020}. 
In one and two dimensions, the $\gamma$ matrices reduce to the Pauli matrices and the basic Hamiltonian $H_s$ involves only two bands.
Our scheme can also be generalized to multiband systems (see Appendix~\ref{App1} for more details).

Here, we briefly review the main results of the topological classification theory based on the concept of BISs. 
More details can be found in Refs.~\cite{Zhang2018,Zhang2019a,Zhang2019b,Zhang2020}.
For a static system $H_s(\bold{k})$, we choose, without loss of generality, a component $h_0(\bold{k})$ to describe the dispersion of $n_d$ decoupled bands.  
The remaining components $h_{i>0}(\bold{k})$ depict the interband coupling and compose a spin-orbit (SO) field 
$\bold{h}_{\rm so}(\bold{k})\equiv(h_1, h_2, \dots, h_d)$. 
If without the SO field, band crossing can occur on $(d-1)$D momentum hypersurfaces (or surfaces for brevity):
\begin{align}~\label{BIS_def}
{\rm BIS}\equiv\{\bold{k}| h_0(\bold{k})=0\}. 
\end{align}
 
The core content of the BIS-based classification is the bulk-surface duality~\cite{Zhang2018}, 
which states that the $d$D bulk topology can be characterized by a $(d - 1)$D topological invariant defined on all BISs:
\begin{equation}~\label{W_BIS}
\mathcal{W}=\sum_{j}\nu_{j},\quad\nu_{j}=\frac{\Gamma(d/2)}{2\pi^{d/2}}\frac{1}{(d-1)!}\int_{{\rm BIS}_j}\hat{\bold{h}}_{\rm so}({\rm d}\hat{\bf h}_{\rm so})^{d-1},
\end{equation}
where $\nu_j$ counts the winding of the SO field on the $j$th BIS.
Here $\Gamma(x)$ is the gamma function,
$\hat{\bold{h}}_{\rm so}=\bold{h}_{\rm so}/|\bold{h}_{\rm so}|$ is the
unit SO field, and $\hat{\bold{h}}_{\rm so}({\rm d}\hat{\bold{h}}_{\rm so})^{d-1}=\epsilon^{i_{1}i_{2}\cdots i_{d}}\hat{h}_{{\rm so},i_{1}}\hat{h}_{{\rm so},i_{2}}\wedge\cdots\wedge\hat{h}_{{\rm so},i_{d}}$, with $\epsilon^{i_{1}i_{2}\cdots i_{d}}$ being the fully antisymmetric tensor and $i_{1, 2,\dots, d}\in \{1, 2,\dots, d\}$. 
This result can also be interpreted from the perspective of topological charges~\cite{Zhang2019a}.
Topological charges are located at the nodes of the SO field, i.e., $\bold{k}=\bold{k}_c$ where $\bold{h}_{\rm so}(\bold{k}_c)=0$.
The bulk topology is classified by the total charges enclosed by BISs, namely,
\begin{equation}~\label{W_charge}
{\cal W}=\sum_{n\in\mathcal{V}_{\rm BIS}}\mathcal{C}_n.
\end{equation}
Here $\mathcal{C}_n$ characterizes the winding of the SO field around the $n$th charge and $\mathcal{V}_{\rm BIS}$ denotes the region enclosed by BISs with $h_0(\bold{k})<0$. 
In the typical case where $\bold{h}_{\rm so}$ is linear near $\bold{k}_c$, the  charge value is simplified as
\begin{equation}
\mathcal{C}_n={\rm sgn}{[J_{\bold{h}_{\rm so}}(\bold{k}_{\rm c})]},
\label{6}
\end{equation}
where $J_{\bold{h}_{\rm so}}(\bold{k})$ $={\rm det}[(\partial h_{{\rm so},i}/\partial k_j)]$  is the Jacobian determinant.

Floquet topological phases can be characterized in a similar way, 
while the major difference is that the applied driving can lead to more BISs, all of which contribute to the bulk topology~\cite{Zhang2020}. 
A Floquet system can be described by an effective Hamiltonian $H_{F}=\ui\ln{U(T)}/T$, where the time-evolution operator $U(T)={\cal T}\exp[-\ui\int_{0}^{t}H(\tau)d\tau]$ with ${\cal T}$ denoting the time ordering~\cite{Eckardt2017_review}. The eigenvalues of $H_{F}$ form the Floquet bands with two inequivalent quasienergy gaps~\cite{Rudner2020_review}. 
For a driven system described by Eq.~\eqref{Ham_total}, the Floquet Hamiltonian also take a Dirac-type form 
\begin{align}~\label{HF_general}
H_{F}(\bold{k})=\sum_{i=0}^{d}h_{F, i}(\bold{k})\gamma_i.
\end{align}
Hence, one can introduce BISs for Floquet systems, determined by $h_{F, 0}(\bold{k})=0$,
and accordingly define the SO field $\bold{h}_{F, {\rm so}}(\bold{k})\equiv(h_{F, 1}, h_{F, 2},\cdots,h_{F, d})$ with which to characterize topological charges. 
Unlike static systems, band crossings of the Floquet bands can appear at both the quasienergy gaps. 
We refer to the gap around the quasienergy $0$ ($\pi /T$) as the $0$ gap ($\pi$ gap) and the BIS living in this gap as the $0$ BIS ($\pi$ BIS).
The topology of the Floquet bands below the $0$ gap is contributed by all $0$ and $\pi$ BISs but with opposite signs~\cite{Zhang2020}:
\begin{equation}~\label{W_Floquet}
\mathcal{W}=\mathcal{W}_0-\mathcal{W}_{\pi}, \quad \mathcal{W}_q = \sum_{j}\nu_{j}^{(q)}.
\end{equation}
Here $\nu_{j}^{(q)}$ represents the topological invariant associated with the $j$th $q$ BIS ($q = 0, \pi$), and 
${\cal W}_{0}$ (${\cal W}_{\pi}$) characterizes the number of boundary modes inside the 0 gap ($\pi$ gap).
This result indicates that BISs play a more fundamental role in classifying topological phases.
Recent work further reveals that beyond identifying the global bulk topology, 
the local topology on each BIS can also have a unique connection to the gapless modes on the boundary~\cite{Zhang2022}.

It is important to note that unlike the classification of static systems where $h_0$ can denote the component in any spin-polarization axis,
topological characterization in Eq.~\eqref{W_Floquet} requires carefully choosing with which component as $h_{F,0}$ to define the BIS;
the chosen component should be one of those dominating the dispersion of Floquet band structure such that all the driving-induced band crossings can be fully reflected in $h_{F,0}({\bf k})$.
However, this constraint does not decrease the flexibility of the characterization by quench dynamics, which we will show below.

\section{Topological characterization by quench dynamics}~\label{secIII}

In this section,  we propose a generic dynamical scheme for the characterization of Floquet topological phases and exemplify it with both 2D and 3D models.
For simplicity, we restrict our discussion to the case that only one component (denoted as $h_0$) dominates the band dispersion.

\subsection{Generic scheme}~\label{scheme}

\begin{figure}
\centering
\includegraphics[width=9cm]{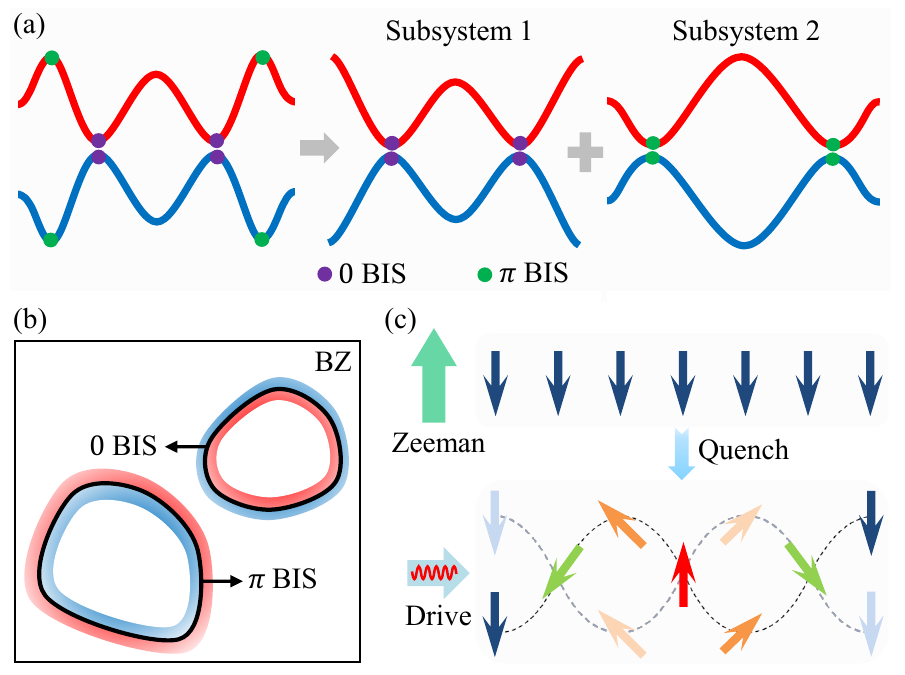}
 \caption{Dynamical characterization scheme.
 (a) Illustration of the basic idea: A Floquet system can be disassembled into multiple static subsystems. Here is a two-subsystem example, with each corresponding to a BIS.
 (b) Illustration of BISs in the BZ, referring to the lower-dimensional subspaces where bands are inverted.
 The colored regions represent the regions of validity for the construction and characterization of effective subsystem Hamiltonians.
 (c) Quench protocol. A large Zeeman field is first added to prepare a fully polarized state, and removed at $t=0$, with a periodic drive being simultaneously applied.
}~\label{Fig1}
\end{figure}

We first show that topological characterization of a Floquet system can be turned into a characterization of multiple static subsystems.
To this end, we adopt the quasienergy operator $Q(t)=H(t)-\ui\partial_t$ to describe the system, 
whose eigenvalues $\varepsilon$ form the Floquet quasienergy bands~\cite{Zhang2020,Rudner2013,Eckardt2015}. 
Under the bases $e^{\ui n\omega t}$ ($n$ is an integer) with $\omega=2\pi/T$ being the driving frequency, a driving field can be written as $V(t)=\sum_{n\neq 0}V^{(n)}e^{\ui n\omega t}$ and the operator $Q(t)$ for the Hamiltonian~\eqref{Ham_total} takes the form
\begin{equation}~\label{Qmatrix}
[Q]_{nn'}=\delta_{nn'}(H_s+n\omega \id)+V^{(n-n')}.
\end{equation}
Here each block $[Q]_{nn'}$ has the same dimension as $H_s$. 
The diagonal blocks $H_s+n\omega\id$ are copies of the static Hamiltonian, 
and the non-diagonal ones $V^{(n-n')}$ couple the $n$th and $n'$th copies. 
The block-matrix form in Eq.~\eqref{Qmatrix} indicates that through exchanging energy with the system, the periodic driving shifts the copies of the static $h_0$ bands in steps of $\omega$, and results in new band crossings (i.e., driving-induced BISs).
Finite driving strength and SO coupling then open gaps on these BISs, rendering the Floquet bands~\cite{Zhang2020}. 
Based on the above analysis, the Floquet system can be treated as a family of subsystems that are periodic in quasienergy, 
with each consisting of $n_d$ bands that are inverted on the corresponding BIS [see Fig.~\ref{Fig1}(a)]. 
Since driving-induced BISs are determined by $h_0({\bf k})=m\omega/2$ ($m=\pm1,\pm2,\cdots$) 
with SO couplings thereon reflecting higher-order effects~\cite{Zhang2022}, 
we will hereinafter refer to the BIS (and also the subsystem) corresponding to a nonzero $m$ as the one of {\it order} $m$.
Due to the bulk-surface duality, the topology on the $j$th BIS must be consistent with the bulk topology 
of the corresponding subsystem of the same order:
\begin{align}~\label{nu_w}
\nu_j=w^{(m)}.
\end{align}
Here the superscript indicates the $j$th BIS is of order $m$ and $w^{(m)}$ is the bulk topological invariant of the corresponding subsystem~\cite{note_bulktopo}.
Together with Eq.~\eqref{W_Floquet}, one can fully characterize Floquet topological phases once the topology of all subsystems is identified.

It should be noted that from the above picture, only the information in the immediate regions adjacent to 
and along BISs is reliable for defining effective Hamiltonians for subsystems [see the colored regions in Fig.~\ref{Fig1}(b)]. 
For example, when the driving takes a special form $V(\bold{k},t)=V_0(\bold{k},t)\gamma_0$,
the effective Hamiltonian for the subsystem of order $m>0$ can be 
obtained by analytic continuation of an effective Hamiltonian defined only {\em on} the corresponding BIS~\cite{Zhang2022}, which reads
\begin{equation}~\label{Heff_sub}
H_{\rm eff}^{(m)}=\left(h_0-\frac{m\omega}{2}\right)\gamma_0+(-1)^m{\cal J}_m\sum_{i>1}h_i\gamma_i,
\end{equation}
where ${\cal J}_m\equiv\sum_{\sum_{j=1}^pn_j=m}\prod_{j=1}^pJ_{n_j}\left(\frac{4V^{(j)}}{j\omega}\right)$,
with $J_n(z)$ being the Bessel function and $n_1,n_2,\dots,n_p\in\{0,1,2,\dots,m\}$.
As a natural result, the characterization should also only use the data measured in the ``regions of validity'' to identify bulk topological invariants $w^{(m)}$,
which serves as one of the basic principles of our dynamical scheme.

We now present how to detect the bulk topology of subsystems by quench dynamics.
We consider a quench along the $\gamma_{\ell}$ axis and write $h_\ell(\bold{k})=m_\ell +u_\ell(\bold{k})$.
Here $u_\ell(\bold{k})$ denotes the momentum-dependent part and $m_\ell$ represents a constant magnetization.
The dynamics is induced by suddenly quenching
a fully polarized trivial phase with $m_\ell\to\infty$ to the target Floquet topological regime at $t=0$ [see Fig.~\ref{Fig1}(c)].
We employ stroboscopic time-averaged spin textures for characterization:
\begin{align}~\label{gammai_average}
\overline{\langle\gamma_i({\bf k})\rangle}_{\ell}=\lim_{N\to\infty}\frac{1}{N}\sum_{n=0}^N \langle\gamma_i({\bf k},t=nT)\rangle_{\ell},
\end{align}
where $\langle\gamma_i({\bf k},t)\rangle_{\ell}={\rm Tr}\big[\rho_{\ell}({\bf k})U^\dagger({\bf k},t)\gamma_iU({\bf k},t)\big]$, 
with the subscript $\ell$ outside the angle brackets denoting the quench axis.
Here, $\rho_\ell$ is the density matrix of the initial state, which satisfies $\gamma_\ell\rho_\ell=-\rho_\ell$.
Since $U(nT)=\cos(nE_FT)-\ui\sin(nE_FT)H_F/E_F$ with $E_F=\sqrt{\sum_{i=0}^d h_{F,i}^{\,2}}$, we have
\begin{align}~\label{gammai_hihl}
\overline{\langle\gamma_i({\bf k})\rangle}_{\ell}=-h_{F,i}({\bf k})h_{F,\ell}({\bf k})/E_F^2({\bf k}){\color{blue},}
\end{align}
which indicates that the characterization $\overline{\langle\gamma_i({\bf k})\rangle}_{\ell}=0$ can determine two different kinds of surfaces, i.e., those momenta where $h_{F,i}({\bf k})=0$ and $h_{F,\ell}({\bf k})=0$, respectively. In particular, the surfaces where $h_{F,\ell}({\bf k})=0$ should appear in all the stroboscopic time-averaged spin textures, independent of the measurement axis $\gamma_i$. Based on this result,
we introduce the concept of {\it dynamical} band-inversion surfaces,
\begin{align}~\label{dBIS_def}
{\rm dBIS}\equiv\left\{\bold{k}\big\vert \overline{\langle\gamma_i({\bf k})\rangle}_{\ell}=0,\forall i\right\},
\end{align}
which identify the surface $h_{F,\ell}({\bf k})=0$ in the deep-quench limit. 
Besides dBISs, one can also find other subspaces where stroboscopic time averages vanish.
We denote
\begin{align}~\label{Lj_def}
L_j\equiv\left\{\bold{k}\big\vert \overline{\langle\gamma_j({\bf k})\rangle}_{\ell}=0,  \overline{\langle\gamma_{i\neq j}({\bf k})\rangle}_{\ell}\neq 0\right\}
\end{align}
to characterize the surfaces $h_{F,j\neq \ell}({\bf k})=0$.

Up to now, there are two special spin-polarization axes that have been mentioned: One is $\gamma_0$ which defines the BIS and the other is the quench axis $\gamma_\ell$.
Thus, the consideration of dynamical characterization should be divided into the following two cases: 
(i) The chosen quench axis is precisely the $\gamma_0$ axis, i.e., $\gamma_\ell=\gamma_0$;
and (ii) they are not the same, namely, $\gamma_\ell\neq\gamma_0$.

{\it Case (i).} In this case, the measured dBISs correspond exactly to the BISs of the Floquet bands.
According to the band structure or by observing how dBISs emerge one by one with the driving frequency~\cite{Zhang2023}, 
one can determine which category each dBIS falls into, such as 0 or $\pi$ type and the order $m$.
Then, a dynamical spin-texture field ${\bf g}({\bf k})=(g_{1},g_{2},\dots,g_{d})$ can be defined to characterize the SO field, with
\begin{align}\label{gi_def}
g_i({\bf k})=-\frac{1}{{\cal N}_k}\partial_{k_\perp}\overline{\langle\gamma_i({\bf k})\rangle}_{\ell}.
\end{align}
Here $k_\perp$ is the momentum perpendicular to dBISs and pointing from the region $h_{F,\ell}({\bf k})<0$ to $h_{F,\ell}({\bf k})>0$, 
and $\mathcal{N}_{k}$ is a normalization factor.
It can be checked that $g_i=h_{F,i}/|h_{F,i}|$ on dBISs.
The subsystem topology can be characterized by the winding of the dynamical field ${\bf g}({\bf k})$ on the corresponding dBISs:
\begin{equation}
w^{(m)}=\frac{\Gamma(d/2)}{2\pi^{d/2}}\frac{1}{(d-1)!}\int_{{\rm dBIS}^{(m)}}\bold{g}(\bold{k})\left[\mathrm{d}\bold{g}(\bold{k})\right]^{d-1},
\end{equation}
where ${\rm dBIS}^{(m)}$ denotes the dBISs that characterize the BISs of order $m$.

{\it Cases (ii).} One needs first to rearrange the $\gamma$ matrices into a new sequence $\{\gamma_{\ell},\gamma_{i_1},\gamma_{i_2},\dots,\gamma_{i_d}\}$ 
which should also satisfy the trace property~\eqref{trace}.
Suppose $\gamma_{i_s}=\gamma_0$ ($1\leq s\leq d$). The surfaces $L_{i_s}$ defined by Eq.~\eqref{Lj_def} (rather than the dBISs) 
are factually the BISs of the Floquet bands,
and can be divided into categories associated with different $m$, denoted as $L_{i_s}^{(m)}$.
To characterize the subsystem topology correctly,
one then needs to construct a spin texture $\langle\gamma^{(m)}_{i_s}({\bf k})\rangle$ for each subsystem
from the measured result $\overline{\langle\gamma_{i_s}({\bf k})\rangle}_{\ell}$.
For a given $m$, the construction of the spin texture $\langle\gamma^{(m)}_{i_s}({\bf k})\rangle$ is as follows: 
The surfaces $L_{i_s}^{(m)}$ together with the dBISs
cut the BZ into patches; the spin polarization of each {\it whole} patch ($\langle\gamma_{i_s}\rangle>0$ or $\langle\gamma_{i_s}\rangle<0$) is determined by 
the measured value of $\overline{\langle\gamma_{i_s}\rangle}_{\ell}$ in the adjacent region of $L_{i_s}^{(m)}$ located in this patch
[i.e., the region of validity as sketched in Fig.~1(b)].
A dynamical field ${\bf g}^{(m)}({\bf k})=(g_{i_1},g_{i_2},\dots,g^{(m)}_{i_s},\dots,g_{i_d})$ 
can then be defined for each subsystem, with the components given by Eq.~\eqref{gi_def} except that
\begin{align}~\label{gis_def}
g_{i_s}^{(m)}({\bf k})=(-1)^{m+1}\frac{1}{{\cal N}_k}\partial_{k_\perp}\langle\gamma^{(m)}_{i_s}({\bf k})\rangle.  
\end{align}
The bulk topology of each subsystem is then characterized by the winding of the corresponding field ${\bf g}^{(m)}({\bf k})$ on all dBISs:
\begin{equation}
w^{(m)}=\frac{\Gamma(d/2)}{2\pi^{d/2}}\frac{1}{(d-1)!}\int_{\rm dBIS}\bold{g}^{(m)}(\bold{k})\left[\mathrm{d}\bold{g}^{(m)}(\bold{k})\right]^{d-1}.
\end{equation}


For both cases, the winding numbers that characterize the topology of quasienergy gaps are obtained by
\begin{align}
\mathcal{W}_0=\sum_{n} w^{(2n)},\quad
\mathcal{W}_{\pi}=\sum_{n}w^{(2n+1)},
\end{align}
where $n$ is an integer.

Before proceeding, we would like to emphasize several points on dBISs: 
(i) Why do we define dBISs as Eq.~\eqref{dBIS_def}? 
The reason is that if we  pick up any other axis but $\gamma_\ell$ to define dBISs, we cannot construct a $d$-component dynamical field to characterize the topology.
From Eq.~\eqref{gammai_hihl}, we see $\overline{\langle\gamma_i({\bf k})\rangle}_\ell\propto-h_{F,i}({\bf k})h_{F,\ell}({\bf k})$ for $i\neq \ell$, which indicates that one can use the measurement $\overline{\langle\gamma_i({\bf k})\rangle}_\ell$ to characterize the corresponding component $h_{F,i}({\bf k})$. However,
when $i=\ell$, it becomes $\overline{\langle\gamma_\ell({\bf k})\rangle}_\ell\propto-h^2_{F,\ell}({\bf k})$, 
which means the full information of $h_{F,\ell}({\bf k})$ cannot be derived from $\overline{\langle\gamma_\ell({\bf k})\rangle}_\ell$.
Thus, to realize the dynamical characterization, 
the only choice is to use $\overline{\langle\gamma_\ell({\bf k})\rangle}_\ell$ to define dBISs, and use other measurements to construct the dynamical field on dBISs.
(ii) The fact that dBISs appear in all the spin textures $\overline{\langle\gamma_i({\bf k})\rangle}_\ell$ can be understood as follows: 
The quench-induced spin oscillation at each ${\bf k}$ corresponds to a spin precession, 
in which the momentum-linked spin rotates about the postquench vector field ${\bf h}_F({\bf k})$.
The dBISs actually refer to the momenta where the spin state is perpendicular to its rotation axis. 
This naturally leads to the conclusion that on a dBIS, the time-averaged spin polarization is zero in any direction.
(iii) dBISs are not the same as BISs, although they may refer to the same momentum subspace in case (i).
While BISs are defined with respect to the (Floquet) band structure [see Eq.~\eqref{BIS_def}], dBISs are defined with respect to dynamical measurements
and used for the dynamical characterization by time-averaged spin textures [Eq.~\eqref{dBIS_def}].
For a Floquet system, the spin-polarization axis $\gamma_0$ used to define BISs needs to be a special one that dominates the band dispersion.
In comparison, the quench axis $\gamma_\ell$ that determines dBISs can be any one in the present dynamical characterization scheme.

\begin{figure}
\centering
\includegraphics[width=9cm]{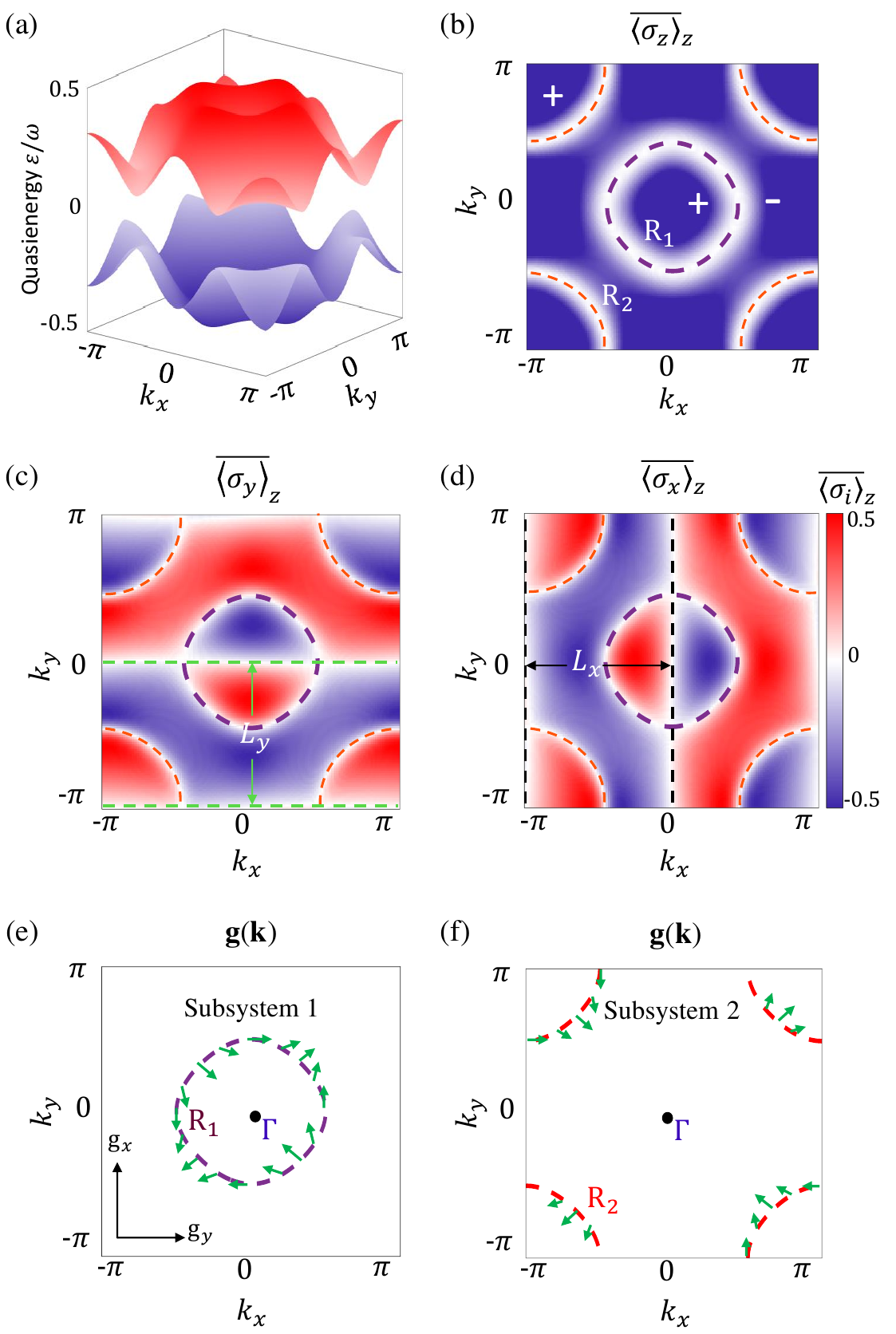}
 \caption{Characterizing the 2D model~\eqref{H2D_def} by quenching $h_z(\bold{k})$. (a) Quasienergy spectrum of the post-quench Hamiltonian in the first BZ with periodic boundary conditions. 
 (b)-(d) Stroboscopic time-averaged spin textures $\overline{\langle \sigma_{i}(\bold{k})\rangle}_z$ ($i=x,y,z$). Two ring-shaped dBISs emerge in all the spin textures: 
 The one labeled as $R_1$ (purple) corresponds to a $\pi$ BIS and the other labeled as $R_2$ (orange) to a $0$ BIS according to the band structure in (a).  
 The symbol $`` + "$ $( `` - ")$ denotes the region where $h_{F, z}(\bold{k})>0$ ($<0$). 
Two dashed lines in $\overline{\langle \sigma_{y}\rangle}_z$ ($\overline{\langle \sigma_{x}\rangle}_z$) with vanishing polarization are denoted as $L_y$ ($L_x$).
 (e) and (f) The dynamical spin-texture field $\bf g(\bold{k})$ (green arrows) is constructed to characterize the topology. 
The post-quench Floquet system is disassembled into two subsystems, whose bulk topology is characterized by the winding of $\bf g(\bold{k})$ on their respective dBISs.
 Here the postquench parameters are $m_x=m_y=0$, $m_z=6t_0$, $t_{\rm so}=t_0$, $\omega=8t_0$ and $V_0=4t_0$.
}~\label{Fig2}
\end{figure}

\subsection{2D model}~\label{2dmodel}

Now we use an experimentally feasible model to illustrate the scheme. We consider the 2D driven model taking the form of Eq.~\eqref{Ham_total} with
\begin{equation}~\label{H2D_def}
H_s(\bold{k})={\bf h}({\bf k})\cdot{\bm \sigma},\quad V(t)=2V_0\cos(\omega t)\sigma_z,
\end{equation}
where $\sigma_{x,y,z}$ are the Pauli matrices and ${\bf h}({\bf k})=(h_x,h_y,h_z)=(m_{x}+2t_{\rm so}\sin k_{x},m_{y}+2t_{\rm so}\sin k_{y}, m_z-2t_0\cos{k_x}-2t_0\cos{k_y})$
depicts the 2D quantum anomalous Hall model that has been realized in optical Raman lattices~\cite{Wu2016,Sun2018a,Liang2023}.
Here $t_0$ $(t_{\rm so})$ represents the spin-conserved (spin-flipped) hopping coefficient and $m_{x,y,z}$ denote the constant magnetization.
The periodic drive can be achieved by modulating the bias magnetic field or the laser frequency~\cite{Zhang2023}. 
For the static Hamiltonian $H_s$,  the bulk topology is simply determined by the Zeeman term $m_z$ ($m_{x,y}=0$): 
the Chern number ${\rm Ch}=0$ for $|m_z|\geqslant4t_0$ and ${\rm Ch}=-{\rm sgn}(m_z)$ for $0<|m_z|<4t_0$. 
The applied drive can, however, largely modify the band structure, leading to a much richer Floquet topological phase diagram~\cite{Zhang2023}.
Figure~\ref{Fig2}(a) displays the quasienergy spectra of a phase with ${\rm Ch}=2$ at $m_z=6t_0$, 
where both the BISs ($R_{1,2}$) are induced by the driving and correspond to $h_z({\bf k})=m\omega/2$
with $m=1$ ($R_{1}$) and $m=2$ ($R_{2}$), respectively  [see Fig.~\ref{Fig2}(b)].
The result in Eq.~\eqref{Heff_sub} gives an expression of $H^{(1)}_{\rm eff}$ ($H^{(2)}_{\rm eff}$), which describes subsystem 1 (2) corresponding to the driving-induced $\pi$ BIS $R_1$ ($0$ BIS $R_2$). 
Note that in Fig.~\ref{Fig2}, the setting $m_z=6t_0$ is comparable to the driving frequency $\omega$, 
indicating that BISs should be defined with respect to the $\sigma_z$ axis, 
for only $h_{F,z}({\bf k})=0$ can reflect all the band crossings in both the 0 gap and the $\pi$ gap.
Otherwise, the topology of the $\pi$ gap cannot be characterized if we use $h_{F,x/y}({\bf k})=0$ to define BISs.

As described in Sec.~\ref{scheme}, the flexibility of the present characterization scheme 
allows us to consider different methods of quenching, which can benefit experimental measurements.
Here we consider both of the aforementioned quench cases: (i) The quench axis is precisely the $\sigma_z$ axis for this 2D driven model,
and (ii) the quench is along another axis. For case (i), the quench is performed by suddenly varying $m_z$ from $m_z\gg t_0$ to $m_z=6t_0$ at time $t=0$, while setting $m_{x,y}=0$.
The periodic driving begins from $t=0$ as well [Fig.~\ref{Fig1}(c)].
We investigate the quench-induced spin dynamics and derive the stroboscopic time-averaged spin textures $\overline{\langle\sigma_{i}(\bold{k})\rangle}_z$ ($i=x,y,z$) based on Eq.~\eqref{gammai_average}. 
The results are shown in Figs.~\ref{Fig2}(b)-\ref{Fig2}(d).
One can see that two ring-shaped structures [i.e., two dBISs as defined in Eq.~\eqref{dBIS_def}] emerge in all the spin textures, 
of which one ($R_1$) corresponds to a $\pi$ BIS and the other ($R_2$) identifies a $0$ BIS.
According to Eq.~\eqref{gi_def}, a dynamical spin-texture field ${\bf g}({\bf k})=(g_{y},g_{x})$ is further constructed and is depicted with green arrows in Figs.~\ref{Fig2}(e) and \ref{Fig2}(f).  
One can find that the dynamical field ${\bf g}({\bf k})$ exhibits nonzero but opposite windings on the two rings~\cite{note_winding},  
giving $\mathcal{W}_{\pi}=w^{(1)}=-1$ and $\mathcal{W}_{0}=w^{(2)}=1$. 
The Chern number of the Floquet bands is ${\rm Ch}=\mathcal{W}_0-\mathcal{W}_{\pi}=2$. 

\begin{figure}
\centering
\includegraphics[width=9cm]{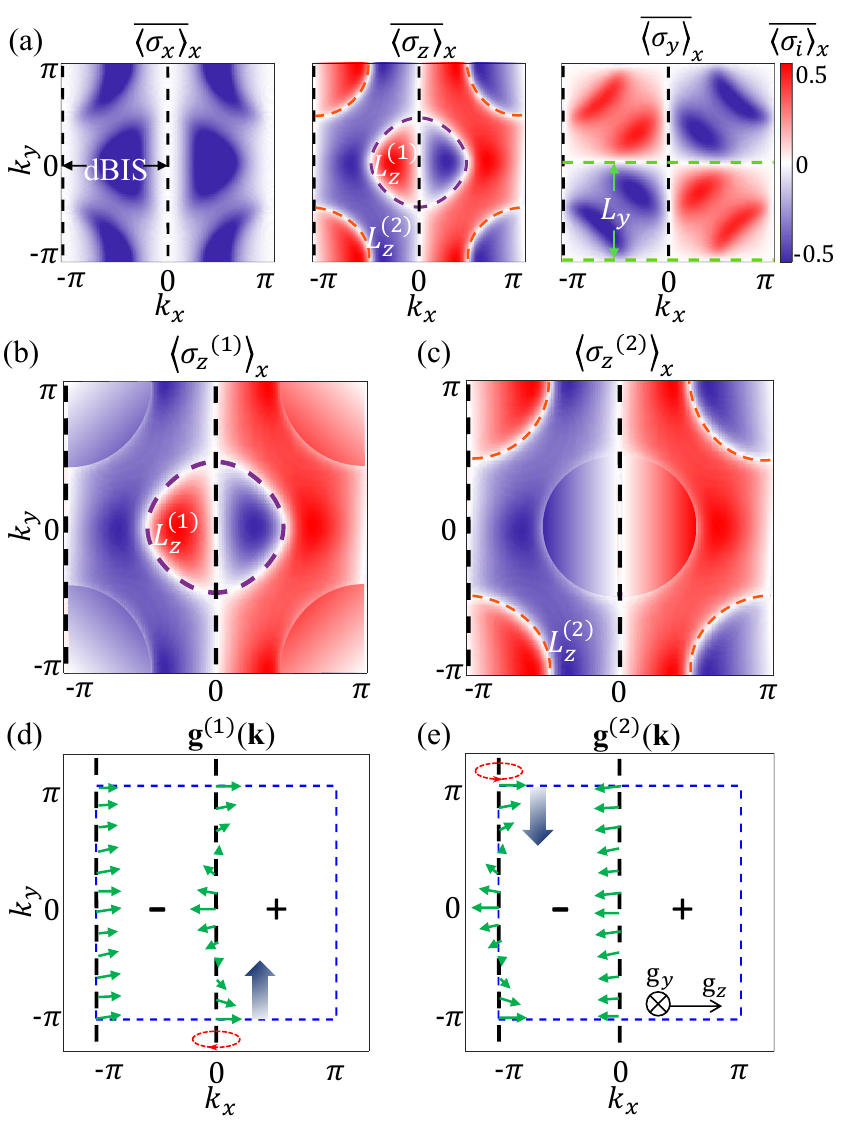}
 \caption{Characterizing the 2D model~\eqref{H2D_def} by quenching $h_x(\bold{k})$. 
(a) Stroboscopic time-averaged spin textures $\overline{\langle \sigma_{i}(\bold{k})\rangle}_x$ ($i=x,y,z$). 
The two lines at $k_x=-\pi,0$ (black) emerge in all the textures and exhibit no polarization, and they are identified as the dBISs. 
In addition, the two rings $L_z^{(1,2)}$ with vanishing polarization only appear in $\overline{\langle \sigma_{z}(\bold{k})\rangle}_x$ and 
the two lines $L_y$ (green) only appear in $\overline{\langle \sigma_{y}(\bold{k})\rangle}_x$.
Here, $L_z^{(1)}$ ($L_z^{(2)}$) corresponds to the BIS with $m=1$ ($m=2$).
(b) and (c) Spin textures $\langle \sigma_{z}^{(1,2)}(\bold{k})\rangle_x$ are drawn from the measured result $\overline{\langle \sigma_{z}(\bold{k})\rangle}_x$ in (a). 
In each texture $\langle \sigma_{z}^{(m)}(\bold{k})\rangle_x$ ($m=1,2$), 
the whole BZ is divided into patches by dBISs and $L_z^{(m)}$.
The spin polarization in each patch is set to have the same sign.
(d) and (e)  The dynamical spin-texture fields ${\bf g}^{(1,2)}(\bold{k})$ are constructed from $\langle \sigma_{z}^{(1,2)}(\bold{k})\rangle_x$ and $\overline{\langle \sigma_{y}(\bold{k})\rangle}_x$    for the two subsystems. 
The nonzero winding of ${\bf g}^{(1)}(\bold{k})$ [${\bf g}^{(2)}(\bold{k})$] along $k_x=0$ ($k_x=-\pi$) characterizes the bulk topology. 
Here, the dashed blue line marks the first BZ, and the sign $`` + "$  $(`` - ")$ denotes the region where $h_{F, x}(\bold{k})>0$ ($<0$). 
The parameters are the same as in Fig.~\ref{Fig2}.}~\label{Fig3}
\end{figure}

For case (ii), we quench the $\sigma_x$ axis by changing $m_x$ from $m_x\gg t_0$ to $m_x=0$, while setting $m_y=0$ and $m_z=6t_0$. 
The other steps are the same as in case (i). 
The stroboscopic time-averaged spin textures $\overline{\langle\sigma_{i}(\bold{k})\rangle}_x$ ($i=x,y,z$) are shown in Fig.~\ref{Fig3}(a).
One sees that two line-shaped structures appear in all the spin textures, corresponding to two open dBISs at $k_x=0$ and $k_x=-\pi$, respectively. 
Other curves with vanishing spin polarization that only appear in $\overline{\langle\sigma_{z}\rangle}_x$ ($\overline{\langle\sigma_{y}\rangle}_x$)
are denoted as $L_z^{(1,2)}$ ($L_y$).
Note that the curve $L_z^{(1)}$ ($L_z^{(2)}$) is in fact the $\pi$ BIS ($0$ BIS).
To characterize the subsystem topology, we need to construct  a spin texture  $\langle \sigma_{z}^{(m)}(\bold{k})\rangle_x$ for each subsystem.
The results are shown in  Figs.~\ref{Fig3}(b) and \ref{Fig3}(c). 
For each $m$, the curves $L_z^{(m)}$ and dBISs cut the BZ into four patches; each patch is painted red or blue according to the spin polarization in the region of validity adjacent to $L_z^{(m)}$ [cf. Fig~\ref{Fig1}(b)].
Based on Eqs.~\eqref{gi_def} and \eqref{gis_def}, dynamical spin-texture fields ${\bf g}^{(1,2)}(\bold{k})=(g_{z},g_{y})$ can be constructed on the dBISs to characterize the two subsystems.
One can see that despite a trivial pattern along the line $k_x=-\pi$ ($k_x=0$), 
the winding of the dynamical field ${\bf g}^{(1)}(\bold{k})$ [${\bf g}^{(2)}(\bold{k})$] on the other dBIS at $k_x=0$ ($k_x=-\pi$) 
characterizes the topological invariant $w^{(1)}=-1$ ($w^{(2)}=1$). The results are the same as case (i), 
which confirms that the present dynamical characterization scheme is independent of the quench axis.

Despite the fact that only one anomalous topological phase is studied here, the dynamical characterization scheme can certainly be applied to 
other 2D Floquet topological phases. In Appendix~\ref{App2}, we show a dynamical characterization of an unconventional topological phase called the anomalous Floquet valley-Hall phase~\cite{Zhang2022,Zhang2023}.


\subsection{3D model}~\label{3dmodel}

\begin{figure}
\centering
\includegraphics[width=9cm]{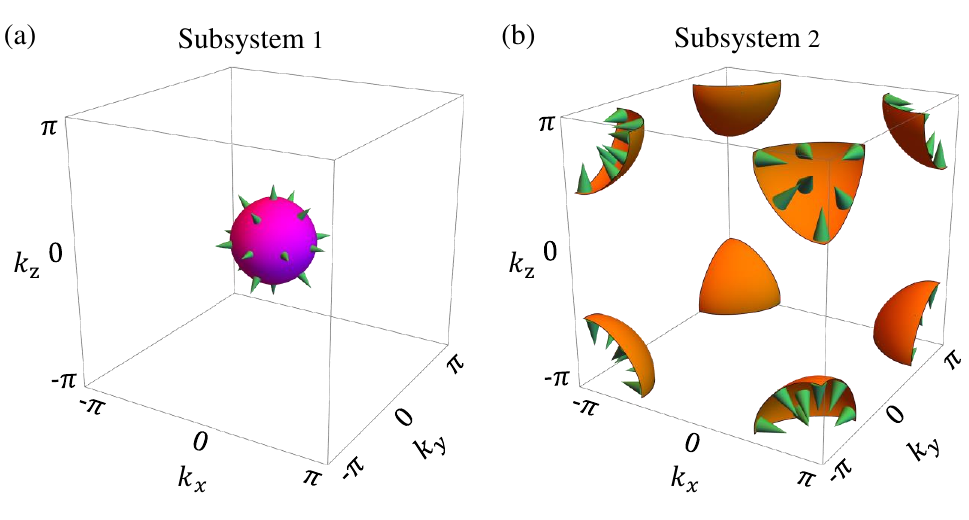}
 \caption{Characterizing the 3D driven model~\eqref{H3D_def}. 
 The whole system is disassembled into two subsystems: One corresponds to the static $0$ BIS (a), and the other corresponds to the driving-induced $\pi$ BIS (b). 
The dynamical spin-texture field ${\bf g}(\bold{k})$ is constructed and plotted as green arrows. 
 Here, we set $t_{\rm so}=t_0$, $m_0=5t_0$, $\omega=18t_0$, and $V_0=2t_0$. More details are given in Appendix~\ref{App3} and Fig.~\ref{Fig8} therein.
 }~\label{Fig4}
\end{figure}

We further apply our scheme to characterize a 3D driven model of the form in Eq.~\eqref{Ham_total} with
\begin{align}~\label{H3D_def}
\begin{split}
h_0(\bold{k})&=m_0-2t_0\sum_{i=1}^3\cos{k_{r_i}},\\ 
h_{i>0}(\bold{k})&=2t_{\rm so}\sin{k_{r_i}}, \, V(t)=2V_0\cos(\omega t)\rm{\gamma}_0.
\end{split}
\end{align}
Here we denote $(r_1,r_2,r_3)\equiv (x,y,z)$ and take ${\rm \gamma}_0=\sigma_z\otimes{\tau}_x$, ${\rm \gamma}_1=\sigma_x\otimes\id$, ${\rm \gamma}_2=\sigma_y\otimes\id$ and ${\rm \gamma}_3=\sigma_z\otimes{\tau}_z$, where $\sigma_{i}$ and ${\tau}_{i}$ are both Pauli matrices. The static Hamiltonian $H_s(\bold{k})=\sum_{i=0}^{3}h_{i}(\bold{k})\gamma_{i}$, having been simulated using solid-state spin systems~\cite{Ji2020,Xin2020}, respects the chiral symmetry $S={\ui}^{2}{\rm \gamma}_0{\rm \gamma}_1{\rm \gamma}_2{\rm \gamma}_3$ and describes 3D chiral topological phases including three regions: (I) $2t_{0}<m_{0}<6t_{0}$ with winding number ${\cal W}=1$; (II) $-2t_{0}<m_{0}<2t_{0}$ with ${\cal W}=-2$; and (III) $-6t_{0}<m_{0}<-2t_{0}$ with ${\cal W}=1$. 
In the presence of the driving $V(t)$, it can be checked that the corresponding Floquet Hamiltonian takes the form of Eq.~\eqref{HF_general}, 
which maintains the same chiral symmetry $S$ and hosts anomalous chiral topological phases that can be identified by BIS characterization theory~\cite{Zhang2022}.

The dynamical characterization of the 3D driven model can be achieved by quenching $m_0$ [belonging to case (i)].
For post-quench parameters $m_0 =5t_0$ and $\omega=18t_0$, 
the Floquet topology of the 3D anomalous chiral topological phase is contributed from two static effective subsystems,  
one corresponding to the static $0$ BIS with $m=0$ [Fig.~\ref{Fig4}(a)] and the other to a driving-induced $\pi$ BIS with $m=1$ [Fig.~\ref{Fig4}(b)]. 
The dynamical spin-texture field ${\bf g}(\bold{k})$ on each surface can be derived from stroboscopic time-averaged spin textures (see Appendix~\ref{App3} for details), 
the winding of which yields $w^{(0)}=w^{(1)}=1$. We then have $\mathcal{W}_{0}=\mathcal{W}_{\pi}=1$, which reveals the nontrivial topology within the two gaps, but ${\cal W}=\mathcal{W}_{0}-\mathcal{W}_{\pi}=0$ indicates that the Floquet bands are topologically trivial.

The numerical results above demonstrate the feasibility of the present dynamical characterization scheme. 
Although only 2D and 3D models are examined, the characterization can easily be generalized to higher dimensions.

\section{The viewpoint of topological charges}~\label{secIV}

In this section, we adopt an alternative viewpoint to examine the dynamical characterization scheme. 
As in static systems~\cite{Zhang2019a,Yi2019}, the dynamical characterization of Floquet topological phases can also be achieved from the viewpoint of topological charges.
Our starting point is still Eqs.~\eqref{dBIS_def} and~\eqref{Lj_def} following the same quench protocol described in Sec.~\ref{scheme}.
Also, two cases need to be considered.

\begin{figure}
\centering
\includegraphics[width=0.5\textwidth]{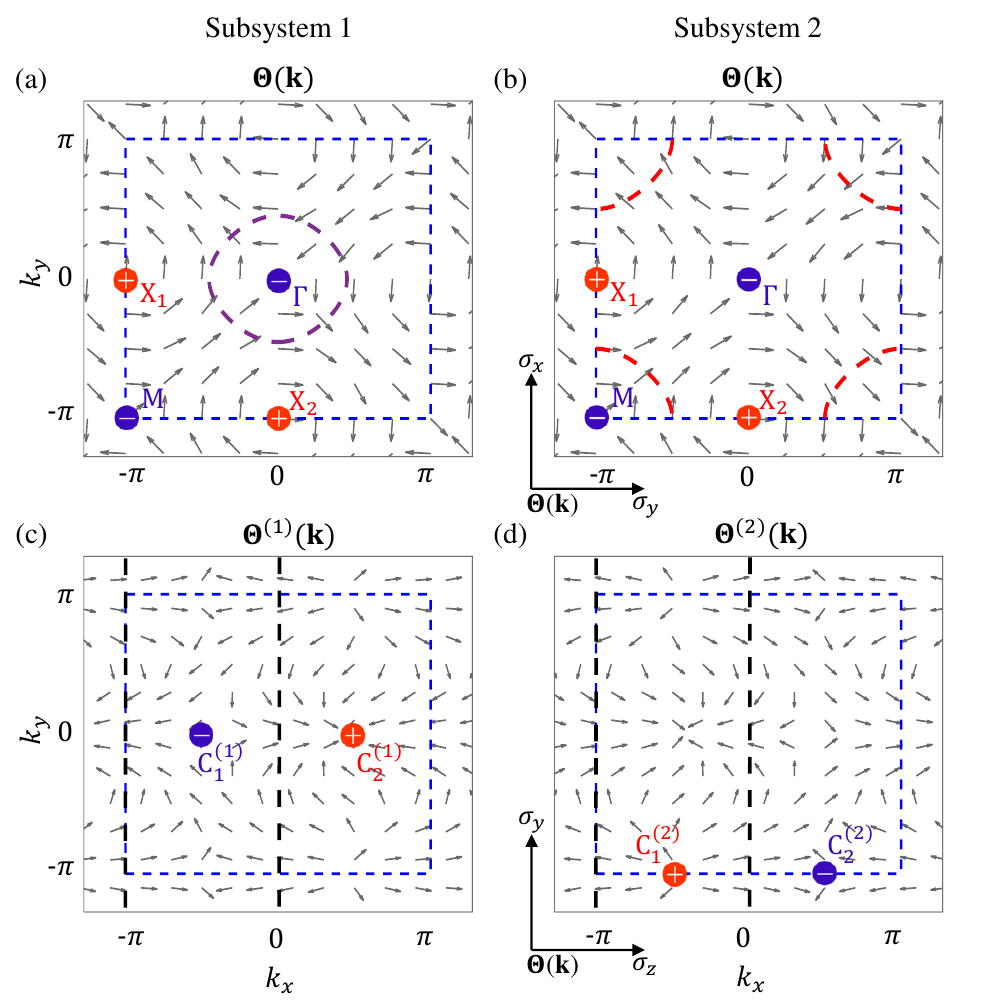}
    \caption{Characterizing the 2D driven model~\eqref{H2D_def} from the viewpoint of topological charges.
    (a) and (b) Characterization by quenching $h_z$. The stroboscopic time-averaged spin textures are shown in Figs.~\ref{Fig2}(b)-\ref{Fig2}(d).
    Both subsystems have four dynamical charges. Their locations are determined by the intersections of $L_{x,y}$.
    The constructed dynamical field $\bold{\Theta(\bold{k})}$ is plotted as arrows, which characterizes the charge value. 
    The corresponding dBIS enclosing only one charge at $\Gamma$ (a) or three charges (b) exhibits different bulk topology.
    (c) and (d) Characterization by quenching $h_x$.
    The stroboscopic time-averaged spin textures are shown in Fig.~\ref{Fig3}(a).
    The two subsystems have different dynamical charges, identified by the intersections of $L_y$ with $L_z^{(1)}$ (subsystem 1) and with $L_z^{(2)}$ (subsystem 2), respectively. Their constructed dynamical fields $\bold{\Theta}^{(1,2)}(\bold{k})$ have opposite $z$-components [cf. Eq.~\eqref{Thetai_m}].
    The same two open dBISs enclosing the charge ${\cal C}_1^{(1)}$ or ${\cal C}_1^{(2)}$ characterizes different bulk topology. 
    Here, the dashed blue line marks the first BZ and the red (blue) dots label dynamical charges with a value $+1$ ($-1$).  The parameters are the same as in Fig.~\ref{Fig2}.}~\label{Fig5}
\end{figure}

For case (i) with $\gamma_\ell=\gamma_0$, the dBISs defined in Eq.~\eqref{dBIS_def} characterize the BISs of the Floquet bands. 
We introduce {\it dynamical} charges that are located at the intersection points ${\bf k}={\bf k}_{\rm dc}$ of all surfaces $L_{j}$ with $j\neq 0$:
\begin{align}
\left\{{\bf k}_{\rm dc}\right\}={\bigcap}_{j=1}^d L_j.
\end{align}
Obviously, in this case, the dynamical charges are exactly the topological charges located at the nodes of the SO field $\bold{h}_{F,{\rm so}}({\bf k})$.
Since $\overline{\langle\gamma_{i}(\bold{k})\rangle}_{\ell}|_{\bold{k}\rightarrow \bold{k}_{\rm dc}}\simeq-h_{F,i}(\bold{k})/h_{F,\ell}(\bold{k}_{\rm dc})$ [see Eq.~\eqref{gammai_hihl}],
we define a dynamical spin-texture field $\bold{\Theta}(\bold{k})=(\Theta_{1},\Theta_{2},\dots,\Theta_{d})$, with its components given by
\begin{equation}~\label{Thetai_def}
\Theta_i(\bold{k})=-\frac{{\rm sgn}[h_{F,\ell}(\bold{k})]}{\mathcal{N}_{k}}\overline{\langle\gamma_{i}(\bold{k})\rangle}_{\ell}.
\end{equation}
The charge value of the $n$th dynamical charge is obtained by
\begin{equation}
\mathcal{C}_n={\rm sgn}{[J_{\bold{\Theta}}(\bold{k}_{\rm dc})]}.
\end{equation}
The characterization of a subsystem of order $m$ reduces to the total dynamical charges enclosed by the corresponding dBISs:
\begin{equation}
w^{(m)}=\sum_{n\in\mathcal{V}_{\rm dBIS}^{(m)}}\mathcal{C}_n,
\end{equation}
where $\mathcal{V}_{\rm dBIS}^{(m)}$ denotes the region surrounded by the corresponding dBISs with $h_0^{(m)}{({\bf k})}<0$~\cite{note_h0}. 
Here, $h_0^{(m)}{({\bf k})}$ denotes the $\gamma_0$-component of the effective Hamiltonian for the subsystem of order $m$.

For case (ii) with $\gamma_\ell\neq\gamma_0$, the $\gamma$ matrices need to be rearranged.
We also suppose $\gamma_{i_s}=\gamma_0$ ($1\leq s\leq d$) in the rearranged sequence,
and denote by $L_{i_s}^{(m)}$ the surface that corresponds to the BIS of order $m$.
We identify the dynamical charges located at the intersection points ${\bf k}={\bf k}^{(m)}_{\rm dc}$:
\begin{align}
\left\{{\bf k}^{(m)}_{\rm dc}\right\}=L_{i_1}\cap L_{i_2}\cap\cdots\cap L_{i_s}^{(m)}\cap\cdots\cap L_{i_d}.
\end{align}
Accordingly, we introduce the dynamical spin-texture field
${\bf \Theta}^{(m)}({\bf k})=(\Theta_{i_1},\Theta_{i_2},\dots,\Theta^{(m)}_{i_s},\dots,\Theta_{i_d})$,
where
\begin{align}~\label{Thetai_m}
\Theta^{(m)}_{i_s}({\bf k})=(-1)^{m+1}\frac{{\rm sgn}[h_{F,\ell}(\bold{k})]}{\mathcal{N}_{k}}\overline{\langle\gamma_{i_s}(\bold{k})\rangle}_{\ell}
\end{align}
and other components are given by Eq.~\eqref{Thetai_def}.
In this case, the dynamical charges do not correspond to the topological charges of the SO field.
The topology of each subsystem is characterized by the corresponding dynamical charges enclosed by dBISs:
\begin{equation}
w^{(m)}=\sum_{n\in\mathcal{V}_{\rm dBIS}}\mathcal{C}_n^{(m)},
\end{equation}
where $\mathcal{V}_{\rm dBIS}$ denotes the region surrounded by dBISs with $h_{\ell}({\bf k})<0$ {\color{blue}~\cite{note_h0}},
and the charge value is given by
\begin{equation}
\mathcal{C}^{(m)}_n={\rm sgn}\left[J_{\bold{\Theta}^{(m)}}\left(\bold{k}^{(m)}_{\rm dc}\right)\right].
\end{equation}

Here we take the 2D driven model~\eqref{H2D_def} as an example to illustrate the dynamical characterization via topological charges. 
The quench process, the chosen parameters, and the calculated stroboscopic time-averaged spin textures are all the same as those in Sec.~\ref{2dmodel}.
We also consider the two quench cases.
In case (i) with $h_z$ being quenched, the lines $L_{x,y}$ in Fig.~\ref{Fig2}(b) intersect at four points, identifying four distinct dynamical charges for both subsystems [see Figs.~\ref{Fig5}(a) and \ref{Fig5}(b)].
The dynamical field ${\bold \Theta}(\bold{k})=(\Theta_y,\Theta_x)$ is constructed from the spin textures according to Eq.~\eqref{Thetai_def}, and determines the charge values.
For subsystem 1 of order $m=1$, only a single charge with ${\cal C}=-1$ at $\Gamma$ point is enclosed by the $\pi$ BIS $R_1$ [Fig.~\ref{Fig5}(a)], 
giving the bulk topological invariant $w^{(1)}=-1$.
For subsystem 2 of order $m=2$, three dynamical charges, one with $C=-1$ at $\Gamma$ and two with $C=+1$ at $X_{1,2}$, are enclosed by the $0$ BIS $R_2$ [Fig.~\ref{Fig5}(b)].
The total charge value gives $w^{(2)}=+1$.
In case (ii), the two subsystems have different dynamical charges  [Figs.~\ref{Fig5}(c) and \ref{Fig5}(d)]. Based on the spin textures in Fig.~\ref{Fig3}(a), two dynamical fields ${\bold \Theta}^{(1,2)}(\bold{k})$ can be constructed, each of which characterizes the charges for the corresponding subsystem.
We find that for subsystem 1 (2), the curves $L_z^{(1)}$ ($L_z^{(2)}$) and $L_y$ have two intersections marking two dynamical charges, and only the charge with $C_1^{(1)}=-1$ ($C_1^{(2)}=+1$) is enclosed by the two open dBISs, which yields $w^{(1)}=-1$ ($w^{(2)}=+1$). 
One can see that in both cases, we have ${\cal W}_{0}=w^{(2)}=1$ and ${\cal W}_{\pi}=w^{(1)}=-1$, consistent with the characterization in Sec.~\ref{2dmodel}.

\section{Shallow Quenches}~\label{secV}

Our results above have shown that by quenching a fully polarized initial state, 
the induced quantum dynamics can characterize the topology of post-quench Floquet Hamiltonian $H_F$.
However, in systems such as ultracold atoms, only finite magnetization can be generated~\cite{Yi2019}, 
so that one needs to consider the case of a shallow quench,  i.e., the quench starts from an initial state that is incompletely polarized.
It has been demonstrated for static systems that the dynamical bulk-surface correspondence 
(and thereby the dynamical characterization) is still valid for a shallow quench, 
since an incompletely polarized initial state can always be transformed to a fully polarized one by a local rotation~\cite{Zhang2019b}. 
Similar reasoning can be applied here.
We shall show that the present dynamical characterization scheme for Floquet systems also works in the incompletely polarized case. 
Some details are given in Appendix~\ref{App4}.

We assume a local rotation $R({\bf k})$ under which the rotated initial state $\rho_\ell'({\bf k})=R({\bf k})\rho_\ell({\bf k})R^\dagger({\bf k})$ becomes fully polarized.
In the rotated frame, the Floquet Hamiltonian is now $H_F'({\bf k})=R({\bf k})H_F({\bf k})R^\dagger({\bf k})$.
Note that $\langle\gamma'_i(nT)\rangle_{\ell}={\rm Tr}\big[\rho'_{\ell}e^{\ui H'_FnT}\gamma'_ie^{-\ui H'_FnT}\big]=\langle\gamma_i(nT)\rangle_{\ell}$,
where $\gamma'_i=R\gamma_iR^\dagger$.
After the rotation, the dynamical characterization becomes the one to characterize the topology of $H_F'({\bf k})$ by a deep quench, and obviously it works.
Since $R({\bf k})$ is local unitary, which is ensured by the fact that $\rho_{\ell}({\bf k})$ and $\rho_{\ell}'({\bf k})$ are in the same trivial
regime, the two Hamiltonians $H_F'({\bf k})$ and $H_F({\bf k})$  must be topologically equivalent. This concludes our proof.

\begin{figure}
\centering
\includegraphics[width=0.5\textwidth]{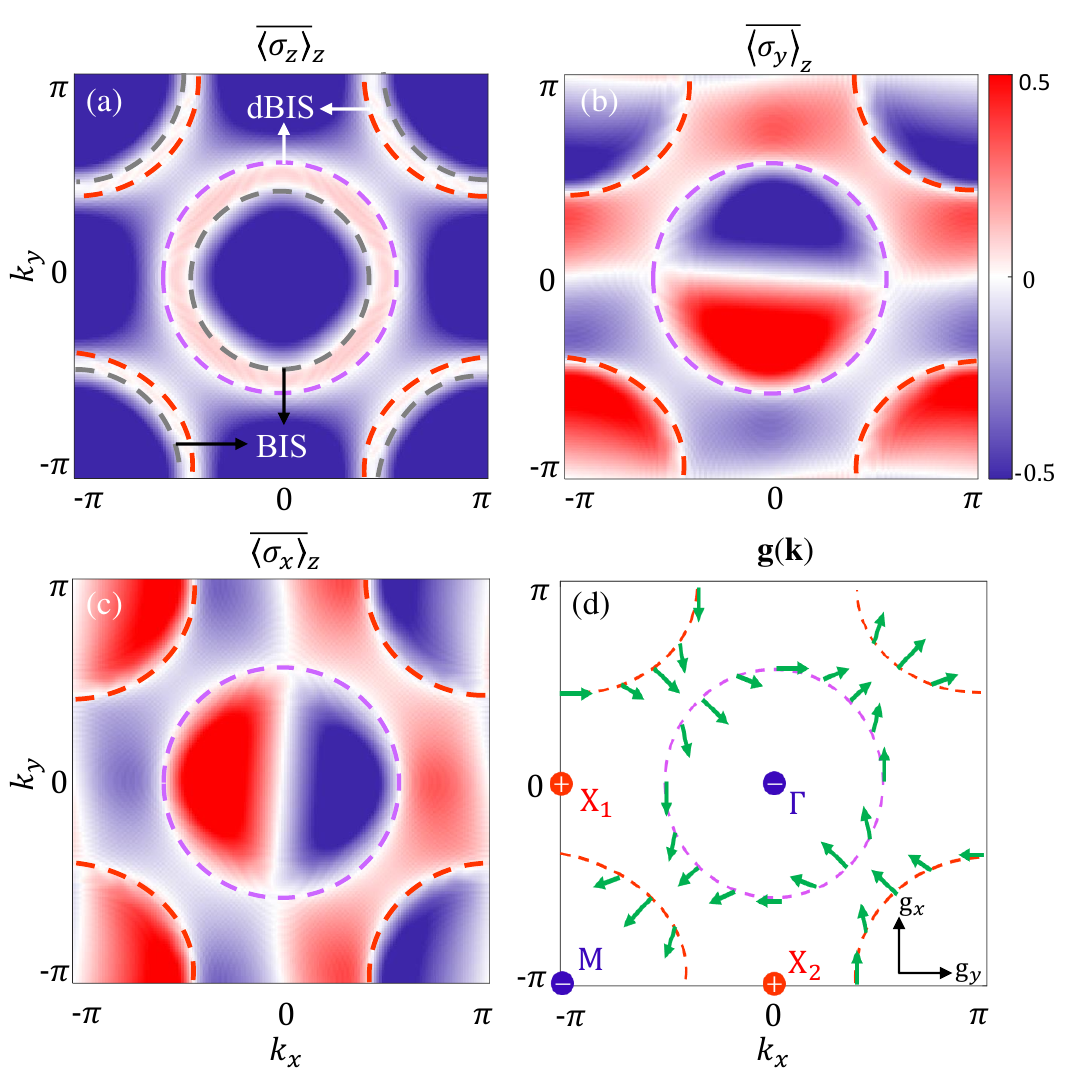}
\caption{Characterizing the 2D driven model~\eqref{H2D_def} by a shallow quench.
[(a)--(c)] Stroboscopic time-averaged spin textures $\overline{\langle \sigma_{i}(\bold{k})\rangle}_z$ ($i=x,y,z$). 
Unlike the deep-quench case in Fig.~\ref{Fig2}, the dBISs do not coincide with the BISs where $h_{F,z}(\bold{k})=0$.
(d) Dynamical characterization by the spin textures in [(a)--(c)].
The winding of the constructed dynamical field $\bf g(\bold{k})$ (green arrows) on dBISs 
and the total dynamical charges (red and blue dots) enclosed by dBISs
yield the same results as the characterization using deep quenches.
Here the postquench parameters are the same as in Fig.~\ref{Fig2}.
}~\label{Fig6}
\end{figure}

For illustration, we still consider the model described by Eq.~\eqref{H2D_def}. 
Here the quench process is realized by changing $m_z$ from a finite value $m_i$ in the trivial regime to the targeted value $m_f=6t_0$.
We investigate the post-quench spin dynamics under the setting $m_i = 5t_0$, with the stroboscopic time-averaged spin textures $\overline{\langle \sigma_{i}(\bold{k})\rangle}_z$ ($i=x,y,z$)  being shown in Fig.~\ref{Fig6}(a)--\ref{Fig6}(c). One can see that two ring-shaped dBISs emerge in all the spin textures.
The dBISs vary with the initial magnetization $m_i$: When $m_i\rightarrow\infty$, they coincide with the BISs where $h_{F,z}(\bold{k})=0$; 
when $m_i\rightarrow4t_0$, the two dBISs deviate from the locations of BISs and move gradually towards each other.
We see that as long as $m_i>4t_0$, each BIS must have its dynamical counterpart, and the emergent topology on dBISs yields a valid characterization of the postquench Floquet topological phase. 
As shown in Fig.~\ref{Fig6}(d), the winding of the dynamical field ${\bf g}(\bold{k})$ on dBISs and the total enclosed charges give the same Floquet topological invariants ${\cal W}_{0}=w^{(2)}=1$ and ${\cal W}_{\pi}=w^{(1)}=-1$ as the deep-quench cases discussed in former sections.

\section{Discussion and Conclusions}~\label{secVI}

We have presented a general and feasible dynamical characterization scheme for a class of generic periodically driven systems 
classified by $\mathbb{Z}$-valued topological invariants.
The scope of application of our proposed dynamical scheme is detailed in Appendix~\ref{App1}.
The present scheme is based on the BIS characterization theory, 
which reduces the characterization of the bulk topology to a sum of contributions of lower-dimensional topology in local momentum subspaces called BISs.
Such a theory has two major advantages: (i) The local topology on BISs can be directly and precisely measured by quench dynamics, which has been experimentally demonstrated in several artificial quantum simulators~\cite{Yi2019,Wang2019,Ji2020,Xin2020,Niu2021,Zhang2023}.
(ii) The classification by local topological structures enables the realization and detection of novel
Floquet topological phases beyond the conventional characterization~\cite{Zhang2022,Zhang2023}.
It has been shown that a particular local topological structure formed in each BIS can 
uniquely correspond to a gapless edge mode, rendering the BIS-boundary correspondence~\cite{Zhang2022}.
Hence, a dynamical scheme based on the BIS characterization potentially has wider applicability.
An example is given in Appendix~\ref{App2} to showcase an unconventional topological phase and its dynamical characterization.

Compared with the previous work in Ref.~\cite{Zhang2020}, the present dynamical scheme is more flexible in performing quenches.
The previous work directly employs the dynamical patterns emerging on BISs to characterize the bulk topology, which depends on how we define the BISs
and thus restricts the quench axis. In comparison, the present scheme introduces effective static Hamiltonians to replace the role of BISs; 
these Hamiltonians have the bulk topology in one-to-one correspondence to the subdimensional topology defined on BISs.
Characterizing a static bulk Hamiltonian can be achieved by a quench along an arbitrary spin-polarization axis~\cite{Zhang2018}.

The present scheme also works under nonideal conditions. Two aspects are discussed here:
(i) Although stroboscopic time-averaged spin textures in Eq.~\eqref{gammai_average} are defined over an infinite interval, 
quench dynamics up to only several oscillation periods $2\pi/\Delta$, where $\Delta$ denotes the local energy gap of the Floquet bands, 
can usually provide sufficient information for topological characterization~\cite{Yi2019,Zhang2023}. 
This indicates that the dynamical measurement is{\color{blue}, to a certain extent,} robust against thermal effects.
(ii) As already discussed in Sec.~\ref{secV}, the proposed dynamical scheme can also employ shallow quenches, 
which loosens the restriction on the preparation of the initial state and has practical benefits for experimental realization.
We have proved that as long as the initial state is in the same trivial regime as the fully polarized state, 
the present scheme always yields a valid dynamical characterization.

Our dynamical characterization theory can also employ a set of quenches with respect to all spin-quantization axes.  
The result in Eq.~\eqref{gammai_hihl} shows an important duality that measuring the $i$th spin component 
after quenching the $\ell$th spin-quantization axis precisely equals to 
measuring the $\ell$th spin component after quenching the $i$th axis. Hence, one can easily check that 
all the stroboscopic time-averaged spin textures that are required for characterization can alternatively be obtained by measuring a single component 
after a set of quenches along all spin-quantization axes~\cite{Zhang2019a,Zhang2019b}. 
However, it should also be noted that whether or not the scheme using a set of shallow quenches can yield a valid characterization of Floquet topological phases is not a straightforward question and deserves further investigation.

In summary, our work provides a multiple-subsystem approach for characterizing $\mathbb{Z}$ Floquet topological phases by quantum quenches.
The present scheme has high flexibility and feasibility in practical applications and can be immediately applied in ultracold atoms or other quantum simulators.

{\it Note added.} Recently, we noted that a dynamical characterization theory has been developed for $\mathbb{Z}_2$ Floquet topological phases based on the concept of 
higher-order BISs~\cite{Zhanglin2023}.

\section*{Acknowledgements}

This work was supported by the National Natural Science Foundation of China (Grants No. 12204187),  
the Innovation Program for Quantum Science and Technology (Grant No. 2021ZD0302000), 
and a startup grant of Huazhong University of Science and Technology (Grant No. 3004012204).

\begin{appendix}

\begin{figure*}
\centering
\includegraphics[width=0.95\textwidth]{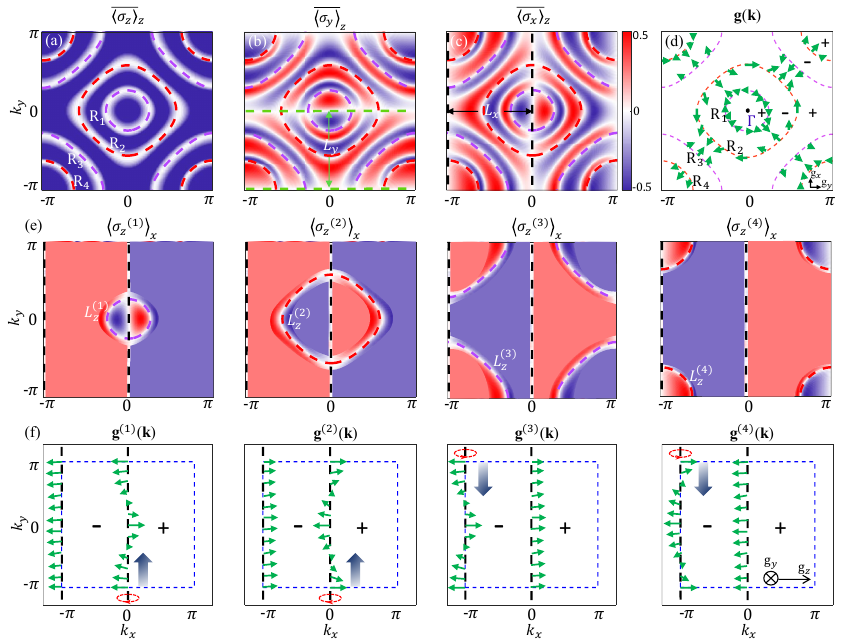}
 \caption{Characterizing the anomalous Floquet valley-Hall phase of the 2D driven model~\eqref{H2D_def}.
 (a)-(d) Characterization by quenching $h_z$.
 Stroboscopic time-averaged spin textures $\overline{\langle \sigma_{i}(\bold{k})\rangle}_z$ ($i=x,y,z$) are shown in (a)-(c).
 Four ring-shaped dBISs emerge (dashed curves) in all the textures: Two correspond to the $\pi$ BISs ($R_{1,3}$) and two correspond to $0$ BISs ($R_{2,4}$).
The dynamical field $\bf g(\bold{k})$ is constructed on the dBISs, which characterizes the topology of subsystems (d). 
Here, $`` + "$ $( `` - ")$ denotes the region where $h_{F,z}(\bold{k})>0$ ($<0$).  
(e) and (f) Characterization by quenching $h_x$. 
Spin textures $\langle \sigma_{z}^{(m)}(\bold{k})\rangle_x$ are derived from $\overline{\langle \sigma_{z}(\bold{k})\rangle}_x$ (e), 
with which the dynamical fields ${\bf g}^{(m)}({\bf k})$ are constructed (f). 
The post-quench parameters are $t_{\rm so}=t_0$, $m_x=m_y=0$, $m_z=5t_0$, $\omega=4t_0$, $V_0=3t_0$, and the initial driving phase $\phi=\pi$.
}~\label{Fig7}
\end{figure*}

\section{The scope of application of our dynamical characterization scheme}\label{App1}

In this appendix, we specify what kinds of Floquet topological phases our dynamic characterization scheme can apply to. 
As shown in Eq.~\eqref{Ham_total}, we consider a class of periodically driven systems realized by applying a periodic drive $V({\bf k},t)$ 
on top of a $d$-dimensional ($d$D) time-independent band structure.
For the latter, we focus on generic gapped topological phases classified by integers 
in the AZ symmetry classes~\cite{AZ1997,Schnyder2008,Kitaev2009,Chiu2016}.
We shall elaborate on this as follows.

First, the basic Hamiltonian for the static system takes the form 
\begin{align}~\label{A1}
H_s(\bold{k})=\bold{h}(\bold{k})\cdot \boldsymbol{\gamma}=\sum_{i=0}^{d}h_{i}(\bold{k})\gamma_{i},
\end{align}
where the $\gamma$ matrices obey the anticommutation relations, and are of dimensionality $n_d =2^{d/2}$ (or $2^{(d+1)/2}$) if $d$ is even (or odd), involving the minimal bands to open a topological gap.
The $\gamma$ matrices can generally be constructed as the tensor product of the Pauli matrices, e.g., in the 3D Hamiltonian~\eqref{H3D_def}.
In one and two dimensions, the Clifford matrices simply reduce to the Pauli matrices [cf. Eq.~\eqref{H2D_def}].
Generally, these $\mathbb{Z}$ topological phases can be topological insulators or superconductors characterized by the winding number (Chern number) in odd (even) dimensions, including classes AIII, BDI, and CII in one dimension, classes A, D, and C in two dimensions, classes AIII, DIII, and CI in three dimensions, and so on. 
In odd dimensions, these $\mathbb{Z}$ topological phases require the protection of the chiral symmetry $S=\ui^{(d+1)/2}\prod_{i=0}^{d}\gamma_i$.
Here, the winding or Chern number characterizes the wrapping number of the map ${\bf n}({\bf k})={\bf h}({\bf k})/|{\bf h}({\bf k})|$ from a $d$D torus $T^d$ to the $d$D spherical surface $S^d$. For such $d$D $\mathbb{Z}$ topological phases described by Eq.~\eqref{A1}, the bulk topology can be characterized by a $(d-1)$D invariant defined on BISs, rendering the bulk-surface duality~\cite{Zhang2018,Zhang2019a,Zhang2019b}.

Second, for the total Hamiltonian, we consider the topology of the Floquet Hamiltonian $H_{F}=\ui\ln {U(T)}/T$. 
Due to the classification of the static Hamiltonian, the resulting Floquet topological phases can also be classified by $\mathbb{Z}$-valued invariants in the ten symmetry classes,
as long as the total Hamiltonian respects the required nonspatial symmetries~\cite{Roy2017,Yao2017}.
Besides, they can also be unconventional topological phases beyond the ten-way classification~\cite{Zhang2022} (see Appendix~\ref{App2} for an example).
When the periodic driving takes the form $V({\bf k},t)=V_{l_1}({\bf k},t)\gamma_{l_1}+V_{l_2}({\bf k},t)\gamma_{l_2}+\cdots$ with $l_i\in\{0,1,\cdots,d\}$, the effective Hamiltonian also takes the basic Dirac-type form shown in Eq.~\eqref{HF_general}. Hence there exist two inequivalent Floquet gaps, named the 0 gap and the $\pi$ gap, respectively, and each gap is characterized by a $\mathbb{Z}$  invariant. A generalized bulk-surface duality has been established~\cite{Zhang2020}: The gap topology can be characterized by a $(d-1)$D invariant defined on the BISs associated with the corresponding gap.

Third, our characterization theory can be generalized to generic multiband systems, for which we assume that each (quasi)energy gap is opened through a group of $n_d$ bands (the minimal requirement).
At each ${\bf k}$, the multiband Hamiltonian can be transformed into a block-diagonal form $H({\bf k})= H_1({\bf k})\oplus H_2({\bf k})\oplus\cdots$~\cite{Zhang2018}. One can see that only those blocks involving $n_d$ crossing bands have nontrivial contribution to the topology. The topology of the bulk bands can then be reduced to topological invariants defined on the corresponding BISs of each block. The contribution of a BIS is effectively determined by a basic Hamiltonian taking the form in Eq.~\eqref{A1}. 
For example, in a 2D three-band system, any two bands that have a band crossing can be locally described by a two-band effective Hamiltonian.
A topological invariant can then be defined on each BIS, which characterizes the bulk topology of the effective Hamiltonian and has a contribution to the Chern numbers of the two involved bands.
Such a generalization applies to both static and Floquet systems. The only difference is that for a Floquet system, one also needs to consider the band crossings resulting from the periodicity of the quasienergy. 

\section{Dynamical characterization of the anomalous Floquet valley-Hall phase}\label{App2}

In this appendix, we apply the present dynamical characterization scheme to more ``unconventional'' topological phases. As demonstrated in Ref.~\cite{Zhang2022}, the concept of BIS provides a systematic way to realize various topological phases by Floquet engineering local topological structures. In particular, when two BISs with opposite topology emerge in the same quasienergy gap, the system may exhibit a topological phase that cannot be classified by conventional topological invariants, e.g., the winding numbers ${\cal W}_{0,\pi}$ and the Chern number, defined for the global topology of the bulk. A typical example is the anomalous Floquet valley-Hall phase~\cite{Zhang2022,Zhang2023}.

For the 2D driven model described by Eq.~\eqref{H2D_def}, the anomalous Floquet valley-Hall phase is marked by four ring-shaped BISs in the first BZ [see Fig.~\ref{Fig7}(a)]: Two correspond to $\pi$ BISs ($R_{1,3}$) and two are $0$ BISs ($R_{2,4}$); they are all induced by periodic driving for the chosen parameters.
Here for clarity of the results, we set an initial driving phase $\phi$ in the driving so that $V(t)=2V_0\cos(\omega t-\phi)\sigma_z$, which does not alter the Floquet topology~\cite{Zhang2020}.
The two types of BISs (two 0 BISs or two $\pi$ BISs) exhibit the same configuration: 
One BIS (e.g., $R_{1}$) surrounds the $\Gamma$ point, and the other (e.g., $R_{3}$)  surrounds $M$.
The dynamical characterization shows that such a configuration results in two opposite contributions to the topological invariant ${\cal W}_{0/\pi}$. 
According to the stroboscopic time-averaged spin textures $\overline{\langle \sigma_{i}(\bold{k})\rangle}_z$ ($i=x,y,z$) in Fig.~\ref{Fig7}(a)--\ref{Fig7}(c), the dynamical field ${\bf g}({\bf k})$  can be constructed [Fig.~\ref{Fig7}(d)], the winding of which on each dBIS gives $w^{(1)}=w^{(2)}=-1$ (for $R_{1,2}$) and $w^{(3)}=w^{(4)}=+1$ (for $R_{3,4}$).
Hence, we have ${\cal W}_{0}=w^{(2)}+w^{(4)}=0$, ${\cal W}_{\pi}=w^{(1)}+w^{(3)}=0$, and the Chern number of the Floquet bands ${\rm Ch}={\cal W}_{0}-{\cal W}_{\pi}=0$.
These topological invariants defined to characterize the global bulk topology are all equal to zero. However, it has been proved that this phase features stable counterpropagating edge states in both quasienergy gaps, and thus transcends the conventional classification~\cite{Zhang2022}.

The characterization can also be performed by quenching the $\sigma_x$ axis. 
By following the same procedure as in Fig.~\ref{Fig3}, we construct four spin textures $\langle \sigma_{z}^{(m)}(\bold{k})\rangle_x$ from the measurement $\overline{\langle \sigma_{z}(\bold{k})\rangle}_x$ [Fig.~\ref{Fig7}(e)], each obtained by extending the data measured in the adjacent region of the corresponding BIS (labeled by $L_z^{(m)}$). 
Accordingly, the dynamical fields ${\bf g}^{(m)}({\bf k})$ are depicted on the two open dBISs [Fig.~\ref{Fig7}(f)], whose winding characterizes the subsystem topology. 
One can easily check that the characterization yields the same results as above: $w^{(1)}=w^{(2)}=-1$ and $w^{(3)}=w^{(4)}=+1$.

\begin{figure*}
\centering
\includegraphics[width=0.95\textwidth]{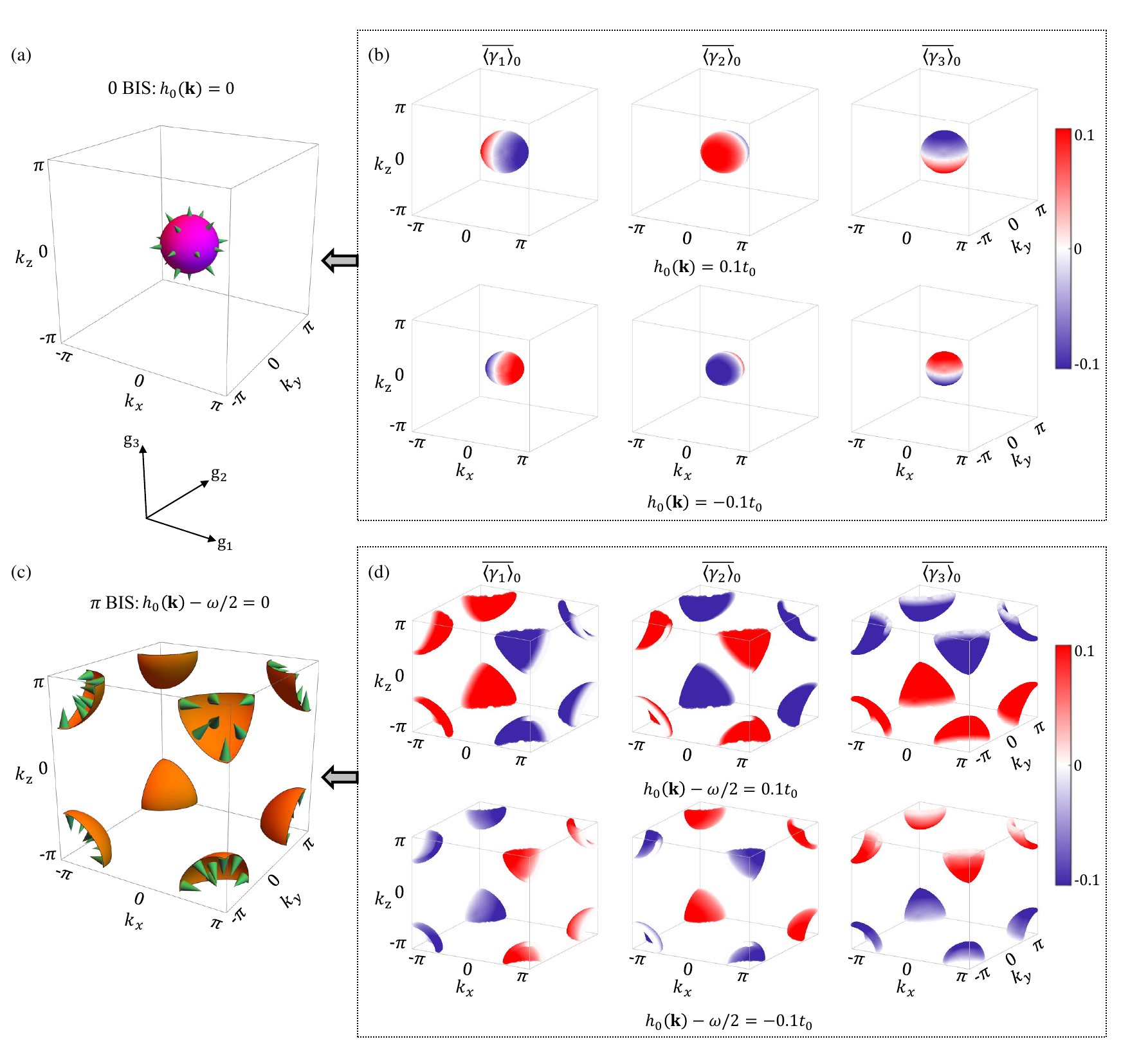}
    \caption{Numerical results of the 3D driven model~\eqref{H3D_def}. 
  The observed stroboscopic time-averaged spin textures $\overline{\langle {\gamma}_i\rangle}_0=0$ ($i=0,1,2,3$) identify a 0 BIS where $h_0({\bf k})=0$ (a) and a $\pi$ BIS where $h_0({\bf k})=\omega/2$ (c). The dynamical field $\bf g(\bold{k})$  (green arrows) is determined by the subtraction of the spin textures on two equal-energy surfaces close to the BIS. For the 0 BIS, the two equal-energy surfaces are $h_0({\bf k})=0.1t_0$ and $h_0({\bf k})=-0.1t_0$ (b). For the $\pi$ BIS, they are $h_0({\bf k})-\omega/2=0.1t_0$ and $h_0({\bf k})-\omega/2=-0.1t_0$ (d). Here the parameters are the same as in Fig. \ref{Fig4}.}~\label{Fig8}
\end{figure*}

\section{Dynamical characterization of 3D topological phases}\label{App3}

In this appendix, we present the details of numerical calculations of the 3D model in Sec.~\ref{3dmodel}.
Here we perform a quench in the $\gamma_0$ axis, while the characterization can also be achieved by other quenches.
As described in the main text, the static Hamiltonian has three topological phase regions distinguished by $m_0$. 
The phase displayed in Fig.~\ref{Fig4} is realized by applying a periodic drive to the phase in region (I).

Similar to the 2D driven model, we detect the 3D Floquet topological phases 
by quenching an initial fully polarized state and deriving stroboscopic time-averaged spin textures $\overline{\langle {\gamma}_i\rangle}_0$ ($i=0,1,2,3$) from quench dynamics.
For the chosen post-quench parameters, two dBISs can be identified by $\overline{\langle {\gamma}_i\rangle}_0=0$ for all $i$ (see Fig.~\ref{Fig8}). 
Since the quench axis $\gamma_0$ is the one that defines the BIS with $h_0({\bf k})=m\omega/2$, 
the two spherical-like surfaces in Fig.~\ref{Fig8} (a) and \ref{Fig8}(c) are in fact the BISs of the Floquet bands: One is a 0 BIS with $m=0$ [Fig.~\ref{Fig8}(a)] 
and the other is a $\pi$ BIS with $m=1$ [Fig.~\ref{Fig8}(b)].
Accordingly, the whole Floquet system is disassembled into two static subsystems, each corresponding to one BIS.

According to Eq.~\eqref{gi_def}, the dynamical spin-texture field ${\bf g}(\bold{k})$ is defined by the gradient of $\overline{\langle {\gamma}_{i>0}\rangle}_0$  perpendicular to the dBISs. Here we show the calculated spin textures on the closed surfaces slightly inside and outside the BISs, respectively [Fig.~\ref{Fig8}(b) and \ref{Fig8}(d)]. 
The direction of the dynamical field ${\bf g}(\bold{k})$ on each BIS can be determined by the subtraction of $\overline{\langle {\gamma}_{i}\rangle}_0$ on the neighboring equal-energy surfaces.
From the winding of ${\bf g}(\bold{k})$, one can obtain the bulk topological invariants for subsystems $w^{(0)}=w^{(1)}=1$, which gives
$\mathcal{W}_{0}=\mathcal{W}_{\pi}=1$, and the winding number of the Floquet bands $\mathcal{W}=\mathcal{W}_{0}-\mathcal{W}_{\pi}=0$.

\section{Some details on shallow quenches}~\label{App4}

Here we give more details on the proof of the shallow-quench method.
Since $U(nT)=\exp\big(-\ui H_F\cdot nT\big)=\cos(E_F\cdot nT)-\ui\sin(E_F\cdot nT)H_F/E_F$, with $E_F=\sqrt{\sum_{i=0}^d h_{F,i}^{\,2}}$,
we have
\begin{align}
\overline{\langle\gamma_i({\bf k})\rangle}_\ell&=\lim_{N\to\infty}\frac{1}{N}\sum_{n=0}^N {\rm Tr}\big[\rho_\ell U^\dagger(nT)\gamma_iU(nT)\big] \nonumber\\
&=\frac{h_{F,i}{\rm Tr}\left[\rho_\ell H_F\right]}{E_F^2},
\end{align}
which yields $\overline{\langle\gamma_i\rangle}_{\ell}=-h_{F,i}h_{F,\ell}/E_F^2$ in a deep-quench case where $\gamma_\ell\rho_\ell=-\rho_\ell$. 
When $\gamma_\ell=\gamma_0$, dBISs defined by Eq.~\eqref{dBIS_def} are exactly the BISs. However, 
for a shallow quench, they are in general not equal [see Fig.~\ref{Fig6}(a)]. 
The definition in Eq.~\eqref{dBIS_def} in fact gives
\begin{align}
{\rm dBIS}=\left\{\bold{k}\big\vert {\rm Tr}\left[\rho_\ell(\bold{k}) H_F(\bold{k})\right]=0\right\},
\end{align}

Now we examine the dynamical characterization.
For shallow quenches, the direction $k_\perp$ is defined to be perpendicular to the contours of the dBISs, i.e., ${\rm Tr}\left[\rho_\ell(\bold{k}) H_F(\bold{k})\right]=0$. 
Suppose that the spin textures are all linear in $k_\perp$ when approaching the dBISs.
The directional derivative on dBISs then reads
\begin{align}\label{dgamma}
\partial_{k_{\perp}}\overline{\langle\gamma_i\rangle}_\ell\simeq-\lim_{k_{\perp}\to0}\frac{1}{2k_{\perp}}\frac{h_{F,i}+{O}(k_{\perp})}
{E_F^{2}+{O}(k_{\perp})}\cdot 2k_{\perp}=-\frac{h_{F,i}}{E_F^{2}}.
\end{align}
We then have $g_i=-\partial_{k_{\perp}}\overline{\langle\gamma_i\rangle}_\ell/{\cal N}_k=h_{F,i}/|h_{F,i}|$. 
As argued in Sec.~\ref{secV}, as long as the incompletely polarized initial state can be connected to the fully polarized one via a local unitary rotation, 
the topological patterns on dBISs should remain the same.
This leads to the conclusion that the dynamical characterization by shallow quenches yields the same result as in the deep-quench case.

\end{appendix}


\begin{thebibliography}{99}
\bibitem{note_bulktopo} According to Eq.~\eqref{W_Floquet}, topological invariants $\nu_j$ should be defined with respect to the Floquet bands below the 0 gap, i.e., to characterize the winding of spin-orbit coupling in the lower bands of $0$ BISs, which are also the upper bands of $\pi$ BISs due to the periodicity of the quasienergy. Therefore, based on Eq.~\eqref{nu_w}, the bulk topological invariant $w^{(m)}$ should be defined to characterize the lower $n_d/2$ bands of the corresponding subsystem if $m$ is even
but to characterize the upper $n_d/2$ bands if $m$ is odd. In our dynamical scheme, we indeed determine $w^{(m)}$ in this way even without particularly mentioning it.
One can certainly define all $w^{(m)}$ in the same manner, which does not change the final result.
\end{thebibliography}
\end{document}